\documentclass{llncs}
\usepackage{hyperref}
\usepackage{graphicx}
\usepackage{booktabs}
\usepackage{svg}
\usepackage{subcaption}
\usepackage{amsmath}
\usepackage{float}
\usepackage{multirow}

\title{A Graph-based Framework for Coverage Analysis in Autonomous Driving}

\author{Thomas Mühlenstädt and Marius Bause}
\date{\today}

\begin{document}

\maketitle

\section{Abstract}

Coverage analysis is essential for validating the safety of autonomous driving systems, yet existing approaches typically assess coverage factors individually or in limited combinations, struggling to capture the complex interactions inherent in traffic scenes.
This paper proposes a graph-based framework for coverage analysis that represents traffic scenes as hierarchical graphs, combining map topology with actor relationships.
The framework introduces a two-phase graph construction algorithm that systematically captures spatial relationships between traffic participants, including leading, following, neighboring, and opposing configurations.

Two complementary coverage analysis methods are presented.
First, a subgraph isomorphism approach matches traffic scenes against a set of manually defined archetype graphs representing common driving scenarios.
Second, a graph embedding approach utilizes Graph Isomorphism Networks with Edge features (GINE) trained via self-supervised contrastive learning to project traffic scenes into a vector space, enabling similarity-based coverage assessment.

The framework is validated on both real-world data from the Argoverse 2.0 dataset and synthetic data from the CARLA simulator.
The subgraph isomorphism method is used to calculate node coverage percentages using predefined archetypes, while the embedding approach reveals meaningful structure in the latent space suitable for clustering and anomaly detection.
The proposed approach offers significant advantages over traditional methods by scaling efficiently to diverse traffic scenarios without requiring scenario-specific handling, and by naturally accommodating varying numbers of actors in a scene.

\section{Introduction}

In autonomous driving, coverage analysis is a crucial step to ensure the safety and reliability of the system.
In most situations, coverage arguments are collected either per coverage factor, or maybe up to 2 or 3 factor interactions.
See for example \cite{foretellix2019} for a production-grade implementation of state-of-the-art coverage analysis.
In contrast to existing approaches, this paper proposes a graph-based framework for coverage analysis.
While graph-based representations of traffic scenes already exist, they have not been specifically designed for this purpose.
This work bridges that gap by utilizing graph-based traffic scene representations explicitly for coverage analysis.

This paper is structured as follows: In chapter \ref{chapter:existing_approaches}, existing coverage and analysis approaches are discussed.
The basis for the following chapters is layed in chapter \ref{chapter:defining_a_traffic_scene_graph}. Afterwards, two coverage approaches 
using the traffic scene graphs are discussed in chapter \ref{chapter:create_subgraphs_for_coverage_analysis} and 
chapter \ref{chapter:implementation_of_graph_embeddings_for_traffic_scene_analysis}.
Finally, the developed methods are applied to Carla and Argoverse 2.0 data in chapter \ref{chapter:application} 
followed by a short summary and outlook. 


\section{Existing coverage and analysis approaches}
\label{chapter:existing_approaches}



There exist a large amount of literature on coverage analysis in the context of autonomous driving.
In the following, some of the most relevant approaches are discussed.

The PEGASUS project (\cite{pegasus2019method}) introduces a systematic, scenario-based methodology for the verification and 
validation of highly automated driving functions, addressing the impracticality of traditional 
distance-based testing approaches. The core of the method is a six-layer model used to 
systematically describe and structure the driving environment, encompassing factors from road geometry to weather conditions. 
This framework is particularly relevant for coverage analysis, as it facilitates a structured decomposition of the vast, 
continuous test space into discrete, manageable "logical scenarios". By systematically 
parameterizing and exploring these logical scenarios across simulation, proving ground and field tests, 
the methodology aims to ensure the completeness of relevant test runs and provides a structured 
foundation for arguing that the automated system has been adequately tested across its entire operational design 
domain (ODD). This shift from random sampling to a structured, scenario-based approach allows 
for a more efficient and comprehensive assessment to generate evidence for a final safety argumentation

In a similar direction, \cite{deGelder2022ontology} focus on defining ontoligies for automated vehicles
in an object oriented manner, with the intention to have a clearly defined and implementable
structures for scenarios, delivering a clear linkage to coverage arguments.

The authors of (\cite{Ries2021traj_clustering}) address the challenge of validating automated driving systems 
in complex urban environments by proposing a trajectory-based clustering method for real-world driving data. 
They extract motion trajectories and associated features from urban driving recordings, apply unsupervised 
clustering to group similar driving behaviours/scenes, then analyze the resulting clusters to reveal common 
scenario types and redundancies. Their approach enables structuring the enormous scenario space, 
supporting more efficient test-set generation and scenario selection for automated vehicle verification. While 
not explicitly focused on coverage analysis, their approach is a good starting point for coverage analysis.

In \cite{Ulbrich2015scene}, the authors identify that key concepts in automated driving—namely scene, 
situation, and scenario—are inconsistently defined in the literature, which complicates the 
development, testing and validation of driving-automation modules. They propose clear 
definitions: a scene is a snapshot of the environment including dynamic and static 
elements plus actors' self-representations; a situation is the set of all circumstances relevant for 
behaviour decision at a given moment, derived from the scene but reflecting the actor's goals and 
values; and a scenario is a temporal development of several scenes in sequence, involving 
actions/events and goals/values. The paper also provides example implementations of these 
definitions in the context of automated-vehicle systems and how they interface with perception, 
planning and control, and testing.

Another important reference is the SAE International recommended practice \cite{ORAD2021taxonomy}, which 
provides many foundational terms in the context of autonomous driving.
It provides a functional taxonomy and clear definitions for key terms related to 
driving automation systems (DAS) used in on-road motor vehicles. The document defines six 
levels of driving automation (Levels 0 through 5) based on the role of three primary actors — 
the human driver, the driving automation system (DAS), and other vehicle systems/components. It introduces 
important related terms such as the Dynamic Driving Task (DDT), Operational Design 
Domain (ODD), Automated Driving System (ADS), and Vehicle Motion Control, 
among others. The 2021-04 revision (superseding the 2018 version) was developed in 
collaboration with ISO/TC 204/WG14 to harmonise global terminology and improve clarity for multi-discipline audiences 
(engineering, legal, media). The document emphasises that it is descriptive (not normative); 
it does not prescribe specifications or impose performance requirements for DAS.

The paper \cite{DBLP:journals/corr/abs-1801-08598} proposes a scenario-based framework for developing and 
validating automated-driving systems across different development phases. It 
introduces three abstraction levels of scenarios—functional, logical, and 
concrete—and discusses how these can be transformed for use in testing. The authors note that 
existing parameter-selection methods, such as equivalence-class or combinatorial testing, lack a 
systematic way to determine meaningful test coverage, highlighting coverage analysis as 
an open challenge in scenario-based validation.

In \cite{Ammann_Offutt_2008}, a standard textbook on software testing although not specifically focused on 
autonomous driving, the authors provide an introduction to software testing. They introduce the concept of coverage and discuss the 
different types of coverage, such as statement coverage, branch coverage, condition coverage, 
and decision coverage. They also discuss the different techniques for measuring coverage, such as 
branch coverage, condition coverage, and decision coverage.

A much more focussed example of coverage analysis in the context of autonomous driving is provided 
by \cite{foretellix2019} in a blog 
post, discussing topics like items, parameters and 
coverage buckets and performance metrics like time to collision and error collections.

The authors of (\cite{wachenfeld2016release}) investigate the significant challenge of safety validation and 
production release for fully autonomous vehicles, positing that established testing concepts are insufficient 
as they fundamentally rely on the human driver's ability to intervene as a safety backup. 
The authors introduce the "approval-trap," a statistical argument demonstrating the unfeasibility of proving 
superior safety through real-world driving, which would necessitate billions of test kilometers. 
This validation gap is presented as a fundamental problem of coverage analysis, where the
technical system must now demonstrably cover the vast operational domain previously managed by the 
human. The paper concludes that overcoming this challenge requires a paradigm shift toward new 
test case generation methodologies, such as critical scenario identification, and the extensive
 use of validated simulation-based tools to achieve sufficient test coverage efficiently.

Well known standards for coverage analysis in the context of autonomous driving are the 
ISO 21448 (\cite{iso21448}) and the UL 4600 (\cite{ul4600}).
The ISO 21448 is a standard for the safety of the intended functionality of road vehicles, 
including automated driving systems.
It defines a framework for coverage analysis, including the definition of coverage criteria and 
the measurement of coverage.
The UL 4600 is a standard for the safety of autonomous products, including automated driving systems.
It defines a framework for coverage analysis, including the definition of coverage criteria and the 
measurement of coverage.

\section{Defining a traffic scene graph}
\label{chapter:defining_a_traffic_scene_graph}






\subsection{Map Graph Construction}
\label{subsec:map_graph}

The map graph serves as the foundational structure for traffic scene graph construction, encoding the 
spatial relationships between lanes in the road network. It is a directed multigraph 
$G_{\text{map}} = (V_{\text{map}}, E_{\text{map}})$ where nodes represent individual lanes and edges 
encode three fundamental spatial relationships between lanes.

Each node $v \in V_{\text{map}}$ represents a lane segment and stores geometric and semantic 
information including:
\begin{itemize}
    \item Lane boundaries (left and right boundaries as polylines)
    \item Lane length
    \item Lane centerline
    \item Intersection status (whether the lane is within an intersection)
    \item Road type and lane type information
\end{itemize}

The map graph defines three edge types that capture different spatial relationships:

\begin{enumerate}
    \item \textbf{Following edges} ($e_{\text{following}}$): Connect lanes that form a continuous 
          path in the same direction. 
    
    \item \textbf{Neighbor edges} ($e_{\text{neighbor}}$): Connect adjacent lanes traveling in 
          the same direction. 
    
    \item \textbf{Opposite edges} ($e_{\text{opposite}}$): Connect lanes traveling in opposite 
          directions. 
\end{enumerate}

The map graph construction algorithm processes the original map data from Argoverse or CARLA  
to extract these relationships. For intersection detection, lanes are analyzed for geometric 
overlap, and all overlapping lanes are marked as intersection lanes to ensure proper handling of 
complex road geometries.

\subsection{Actor Graph Construction}
\label{subsec:actor_graph}

The actor graph $G_{\text{actor}} = (V_{\text{actor}}, E_{\text{actor}})$ represents the dynamic 
relationships between vehicles and other actors in a traffic scene at a specific timestep. This 
graph is constructed using a two-phase algorithm that separates relation discovery from graph 
construction. The second step serves to reduce the number of relations in the graph for computational tractability, 
and enables the representation of relationships between actors through indirect paths involving multiple intermediate nodes.

Each node $v \in V_{\text{actor}}$ represents an actor (vehicle, pedestrian, etc.) and stores the 
following attributes:

\begin{itemize}
    \item \textbf{Primary lane ID}: The lane on which the actor is primarily located
    \item \textbf{Lane IDs}: List of all lanes the actor occupies (for actors spanning multiple 
          lanes)
    \item \textbf{Longitudinal position} ($s$): Position along the primary lane's centerline
    \item \textbf{3D position} ($x, y, z$): Cartesian coordinates
    \item \textbf{Longitudinal speed}: Speed along the lane direction
    \item \textbf{Actor type}: Classification (vehicle, pedestrian, etc.)
    \item \textbf{Lane change indicator}: Boolean flag indicating if the actor changed lanes from 
          the previous timestep
\end{itemize}

Note that the node structure is extensible---users can add additional attributes such as 
acceleration, heading angle, or vehicle dimensions as needed for their coverage model.

Each edge $e \in E_{\text{actor}}$ represents a relationship between two actors and contains:

\begin{itemize}
    \item \textbf{Edge type}: One of four relation types between actors (leading, following, neighbor, opposite)
    \item \textbf{Path length}: Distance in meters along the lane network 
\end{itemize}

Similar to nodes, edge attributes can be extended with additional information such as relative 
velocities, time-to-collision, or interaction probabilities.

\label{subsubsec:relation_types}

The actor graph defines four types of relationships between actors, ordered by semantic hierarchy:

\begin{enumerate}
    \item \textbf{Following/Leading} (\texttt{following\_lead}, \texttt{leading\_vehicle}): 
          Represents longitudinal relationships where one actor follows another in the same 
          direction and lane.  
    
    \item \textbf{Neighbor} (\texttt{neighbor\_vehicle}): Represents lateral relationships 
          between actors on lanes traveling in the same direction. The actors need not be on 
          immediately adjacent lanes.
    
    \item \textbf{Opposite} (\texttt{opposite\_vehicle}): Represents relationships between actors 
          on lanes traveling in opposite directions. Similar to neighbor relationships, the actors 
          need not be on immediately opposite lanes.
\end{enumerate}

\begin{figure}[H]
      \centering
      \begin{subfigure}[b]{0.48\textwidth}
          \centering
          \includegraphics[width=\textwidth]{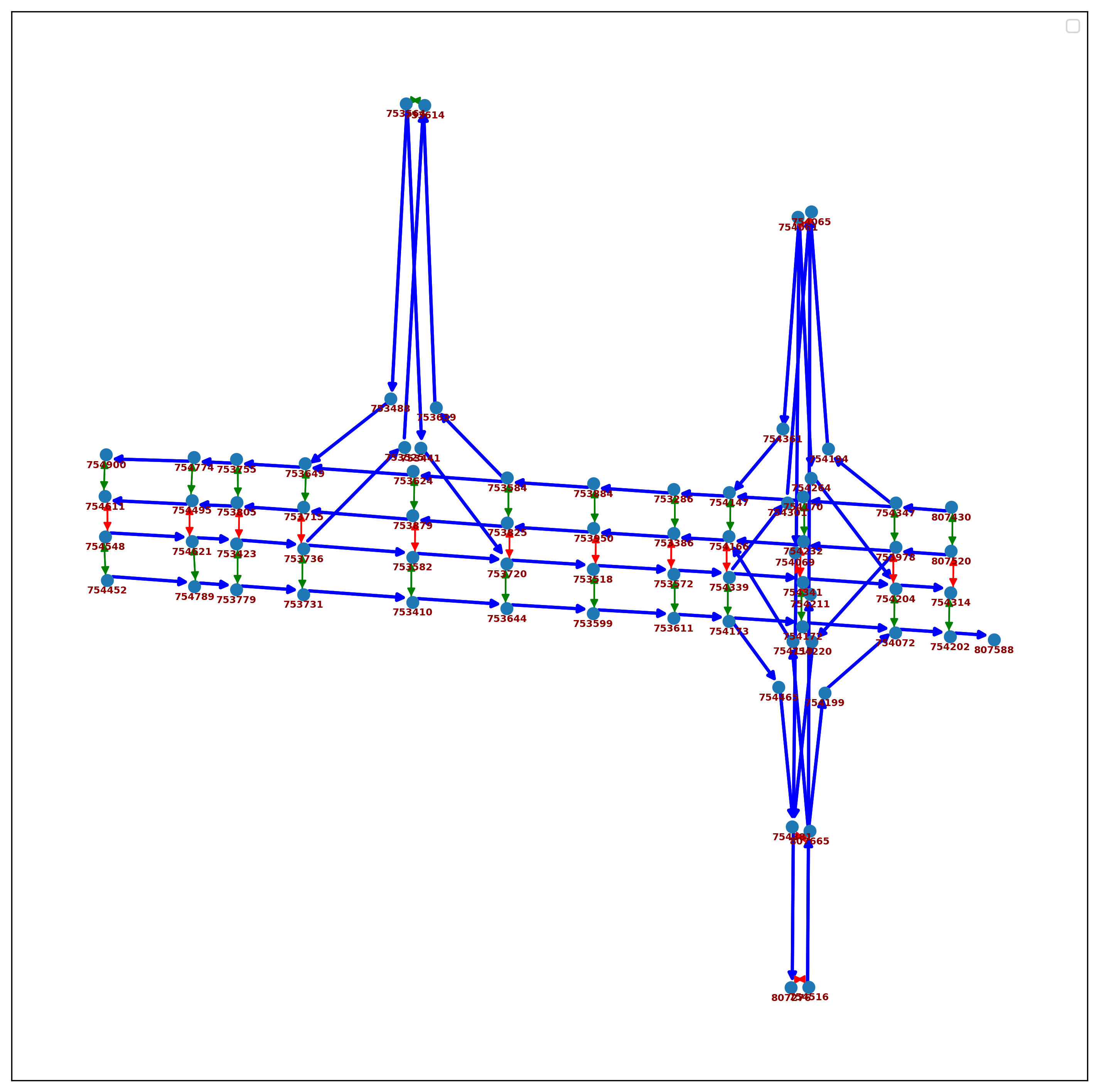}
          \caption{Example graph representation of a map.}
          \label{fig:map_graph_representation}
      \end{subfigure}
      \hfill
      \begin{subfigure}[b]{0.48\textwidth}
          \centering
          \includegraphics[width=\textwidth]{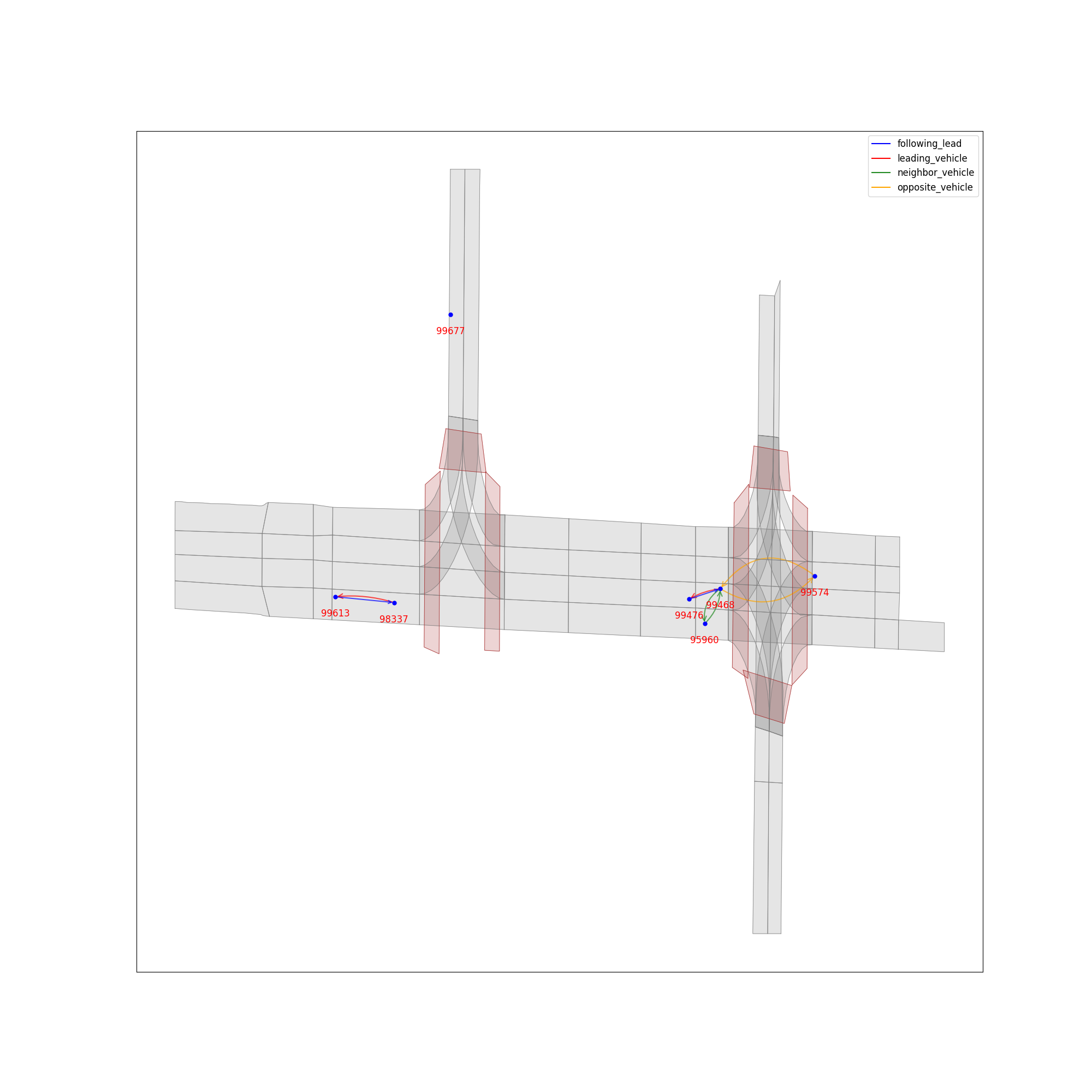}
          \caption{Example traffic graph.}
          \label{fig:traffic_scene_timestep}
      \end{subfigure}
      \caption{The map graph (left) serves as the foundational structure for determining actor relations, encoding spatial relationships between lanes. 
      The traffic scene (right) shows actors and their relationships at a specific timestep. 
      Note that actors can be disconnected if they are far enough apart, as the graph construction algorithm only creates edges between actors within the specified distance thresholds.}
      \label{fig:map_and_scene_graphs}
  \end{figure}

\subsection{Hierarchical Graph Construction Algorithm}
\label{subsec:construction_algorithm}

The actor graph construction follows a two-phase approach that separates relation discovery from 
graph building, enabling hierarchical processing and preventing redundant edge creation.
\begin{table}[H]
      \centering
      \caption{Input parameters for actor graph construction}
      \label{tab:construction_parameters}
      \begin{tabular}{llp{6cm}l}
      \hline
      \textbf{Parameter} & \textbf{Type} & \textbf{Description} & \textbf{Value} \\
      \hline
      \multicolumn{4}{l}{\textit{Distance limits (discovery phase)}} \\
      \hline
      \texttt{max\_distance\_lead\_veh\_m} & float & Maximum distance in meters for leading/following 
      relationships & 100 \\
      \texttt{max\_distance\_neighbor\_forward\_m} & float & Maximum distance in meters for forward 
      neighbor relationships & 50 \\
      \texttt{max\_distance\_neighbor\_backward\_m} & float & Maximum distance in meters for backward 
      neighbor relationships & 50 \\
      \texttt{max\_distance\_opposite\_forward\_m} & float & Maximum distance in meters for forward 
      opposite relationships & 100 \\
      \texttt{max\_distance\_opposite\_backward\_m} & float & Maximum distance in meters for backward 
      opposite relationships & 10 \\
      \hline
      \multicolumn{4}{l}{\textit{Node distance limits (construction phase)}} \\
      \hline
      \texttt{max\_node\_distance\_leading} & int & Maximum number of edges in actor graph path for 
      leading/following & 3 \\
      \texttt{max\_node\_distance\_neighbor} & int & Maximum number of edges in actor graph path for 
      neighbor & 2 \\
      \texttt{max\_node\_distance\_opposite} & int & Maximum number of edges in actor graph path for 
      opposite & 2 \\
      \hline
      \multicolumn{4}{l}{\textit{Timestep configuration}} \\
      \hline
      \texttt{delta\_timestep\_s} & float & Time step increment in seconds for temporal graph 
      creation & 1.0 \\
      \hline
      \end{tabular}
      \end{table}

\subsubsection{Phase 1: Relation Discovery}
\label{subsubsec:discovery_phase}

The discovery phase identifies all potential relationships between actor pairs based on distance 
constraints and map graph structure. For each pair of actors $(A, B)$ at timestep $t$, the 
algorithm determines their primary lanes and checks if a path exists in the map graph connecting 
them. The path structure determines the relationship type:

\begin{itemize}
    \item \textbf{Following/Leading}: If both actors are on the same lane, or if the connecting lane path consists 
          entirely of following edges.
    
    \item \textbf{Neighbor}: If the connecting lane path contains exactly one neighbor edge and the remaining edges 
          are following edges. Note that this does not require the actors to be on immediately 
          adjacent lanes. 
    
    \item \textbf{Opposite}: If the path contains exactly one opposite edge and the remaining edges 
          are following edges. Similar to neighbor relationships, this does not require the actors 
          to be on immediately opposite lanes.
\end{itemize}

Far distance actors are not considered for relation discovery. This is done using two distance checks: 
the lane-based path length ensures that 
curved roads are properly accounted for, while the Euclidean distance ensures that only actors in 
close proximity are considered. The maximum distances are defined in Table~\ref{tab:construction_parameters}. 
All discovered relationships are stored in a relations dictionary 
for the construction phase.

\subsubsection{Phase 2: Hierarchical Graph Construction}
\label{subsubsec:construction_phase}

The construction phase builds the actor graph by adding edges in a hierarchical order, ensuring 
that higher-priority relationships are established first and redundant edges are prevented.

\paragraph{Construction Order}

Edges are added in three stages, following the semantic hierarchy:

\begin{enumerate}
    \item \textbf{Leading/Following edges} (highest priority): All following/leading relationships 
          are processed first, sorted by path length (shortest first).
    
    \item \textbf{Neighbor edges} (medium priority): Neighbor relationships are processed second, 
          sorted by absolute path length.
    
    \item \textbf{Opposite edges} (lowest priority): Opposite relationships are processed last, 
          sorted by absolute path length.
\end{enumerate}

\paragraph{Redundancy Prevention}

\begin{figure}[H]
    \centering
    \begin{subfigure}[b]{0.48\textwidth}
        \centering
        \includegraphics[width=\textwidth]{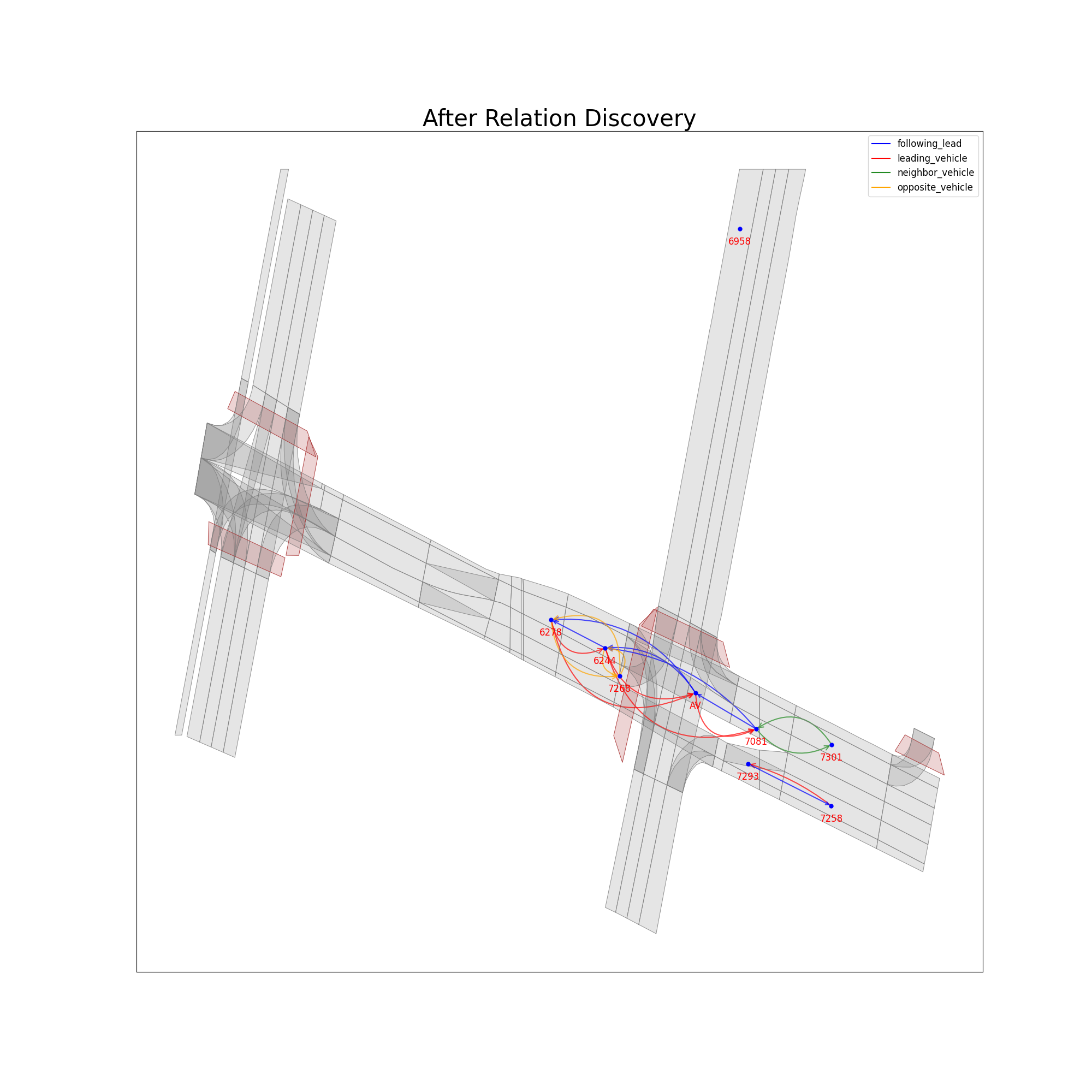}
        \caption{After relation discovery}
        \label{fig:graph_after_discovery}
    \end{subfigure}
    \hfill
    \begin{subfigure}[b]{0.48\textwidth}
        \centering
        \includegraphics[width=\textwidth]{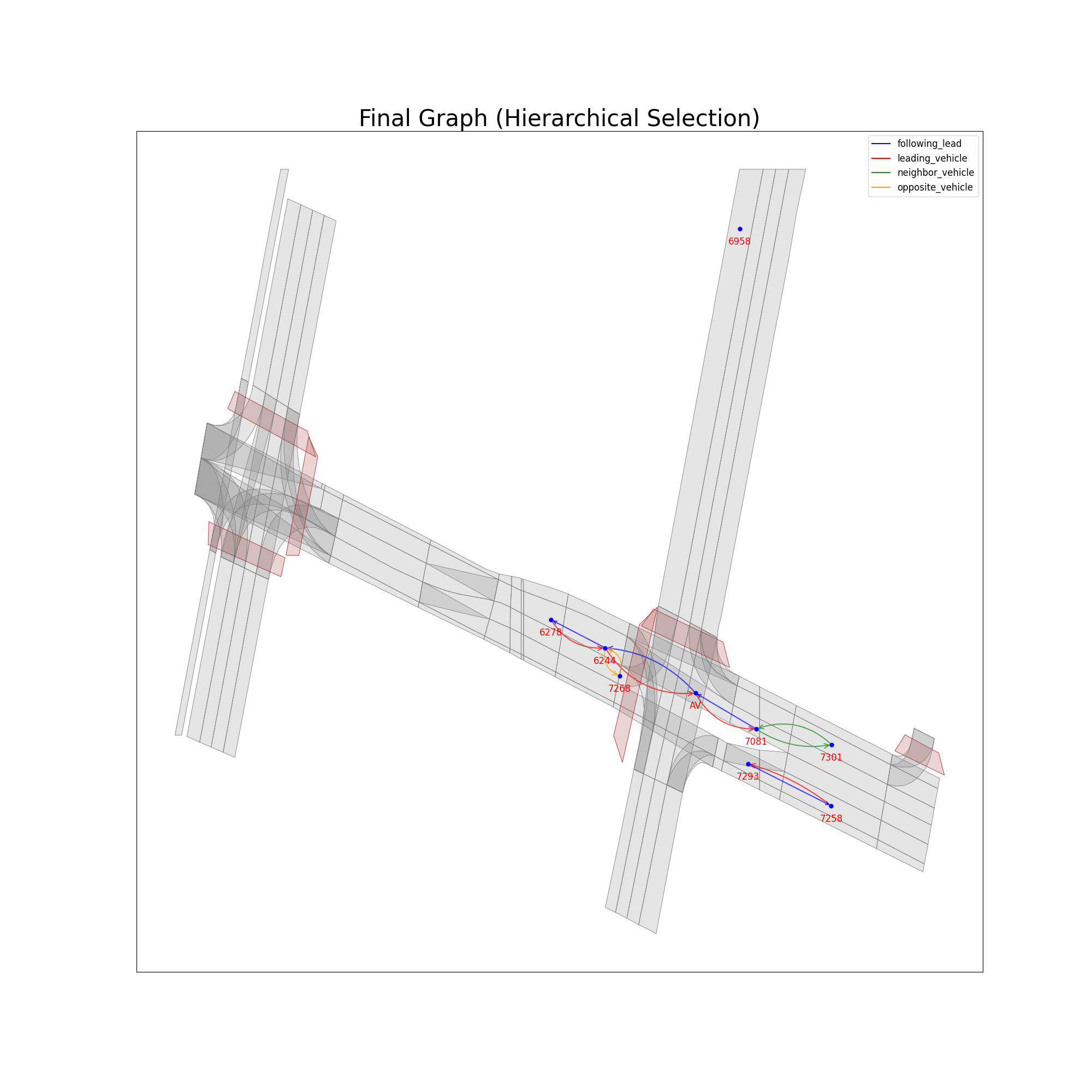}
        \caption{Final graph after hierarchical selection}
        \label{fig:graph_final}
    \end{subfigure}
    \caption{Comparison of the actor graph after relation discovery and after hierarchical 
    selection with redundancy prevention. The discovery graph contains all relations found within 
    the distance limits, while the final graph removes redundant edges that can be represented 
    through existing paths, significantly reducing the number of edges while preserving connectivity.}
    \label{fig:graph_construction_comparison}
\end{figure}

Before adding an edge between actors $A$ and $B$, the algorithm checks if a path already exists 
in the current actor graph with length $\leq \text{max\_node\_distance}$ for that relation type (see Table~\ref{tab:construction_parameters}). 
This prevents redundant edges that would create triangles or long direct paths when an indirect path is already encoded in the graph 
(see Figure~\ref{fig:graph_construction_comparison}).

For leading/following and neighbor relationships, each direction ($A \to B$ and $B \to A$) is 
checked independently, as these relationships can be asymmetric. For opposite relationships, if 
either direction has an existing path, both directions are rejected, as opposite relationships are 
inherently symmetric.

The path checking uses breadth-first search to find the shortest path in the actor graph, counting 
the number of edges (not nodes) in the path. This ensures that if a path exists with $k$ edges 
where $k \leq \text{max\_node\_distance}$, the direct edge is not added.

The graph is updated immediately after each edge addition, ensuring that path checks for 
subsequent edges use the current graph state. This incremental approach guarantees that the 
hierarchical construction works correctly and prevents race conditions.

Figure~\ref{fig:graph_construction_comparison} illustrates the effect of redundancy prevention 
by comparing the graph after relation discovery with the final graph after hierarchical selection. 
The discovery graph respects the maximum distances chosen for discovery, but still contains 
redundant relations. For example, three vehicles in a row can be described by pairwise 
lead-follow relations, but the relation of the last vehicle to the first vehicle is already 
encoded in the graph through the intermediate vehicle. Similarly, for a row of vehicles to a 
neighboring or opposite vehicle, we only need the opposite/following relation to the closest 
vehicle in the row; the other actor relations are encoded in the graph. This redundancy prevention 
reduces the number of edges in the graph significantly, making it more efficient while preserving 
all necessary connectivity information.

\subsubsection{Input Parameters}
\label{subsubsec:parameters}

Table~\ref{tab:construction_parameters} summarizes all input parameters for the actor graph 
construction algorithm, their descriptions, and the values used in our experiments.

The distance limits in the discovery phase determine which potential relationships are identified, 
while the node distance limits in the construction phase control redundancy prevention. The 
asymmetric limits for opposite relationships (100m forward vs. 10m backward) reflect that vehicles 
traveling in opposite directions are more relevant when they are ahead rather than behind.

All parameters are configurable and can be adjusted based on the specific application 
requirements, road network characteristics, or desired graph density. The values shown in 
Table~\ref{tab:construction_parameters} represent default values that work well for urban traffic 
scenarios, but may need adjustment for highway scenarios (longer distances) or dense urban 
intersections (shorter distances).


\subsection{Time based graph representations}
The graph creation strategy described above is assuming to create a graph for a single timestep. 
One possible extension is to create a graph which contains information for multiple timesteps. In 
order to create a time indexed graph, first the actor graphs for all timesteps are created. The 
joint set of all nodes across all timesteps is then determined. Node attributes of single timestep 
graphs are combined into a tensor of node attributes, introducing a new dimension for the time 
index. While the nodes are somehow time independent by this strategy, the edges are modeled as 
time dependent. Here it is utilized, that there can be multiple edges between the same pair of 
nodes, keeping potential edge attributes and adding an additional edge attribute to indicate the 
time stamp of the edge.

This extension allows the graph to easily represent changing states, e.g. the relationship between 
two vehicles changes from neighbor to leading vehicle, i.e. a cut in happened. One important 
special case for this type of processing is to handle actors, which are not present in all time 
steps, i.e. they appear and disappear during the time window. Especially if node attributes are 
used in the processing, a suitable imputation strategy is necessary to avoid non-equal dimensions 
of tensors for node attributes.

The methods derived in the following chapters can be applied to this extended graph representation 
as well, even though the focus in this paper will be on single timestep graphs.

\section{Create subgraphs for coverage analysis}
\label{chapter:create_subgraphs_for_coverage_analysis}

There is a lot of knowledge in the literature on how to define archetypes of traffic scenes, like e.g. with lead vehicle or opposite traffic vehicles in different circumstances.
Once an archetype is defined and expressed as graph, a special property of graphs can be used. 
Two graphs are isomorphic if they have the same structure, regardless of the node and edge labels.
As the archetypes are not necessarily involving a lot of actors, these are more like subsets of actual traffic scenes.
A very simple example might be 2 vehicles on the same lane, driving in the same direction and another vehicle driving on a neighboring lane.
This situation can be represented by a graph with 3 nodes and 2 edges. 
In most real traffic situations however, there will be additional actors present, so that we are not searching for isomorphic graphs, but rather want to check if any subgraph of $G$ is isomorphic to the archetype graph $A$.
This is an example of a subgraph isomorphism problem.
While this problem is NP-hard, the graphs considered here are rather small, so the computational time is reasonable.
One such algorithm is the VF2 algorithm, which is implemented in the NetworkX library (see \cite{cordella2004subgraph}).
The strategy we are then applying is the following:

\begin{enumerate}
    \item Define a set of subgraphs $S$ that are considered to be archetypes of traffic scenes, e.g. unprotected left turns with opposite traffic or lead vehicle following situations.
    \item Define which node and edge attributes are considered for the isomorphism check.
    \item Create an empty dataframe $C$ with a column for each subgraph in $S$
    \item Define the set of traffice scenes (e.g. from Carla or Argoverse) defined as graphs $G$
    \item For each graph $G$, check if any subgraph of $G$ is isomorphic to any subgraph in $S$ and note the result in a new row in table $C$
\end{enumerate}

This strategy can be described to some degree as a bottom up approach: Starting from a detail level, individual situations are defined.
Then going upwards to different datasets, it is checked, if the archetype is present.
Also, follow up analysis of the created coverage dataframe can be performed. For example,

\begin{itemize}
    \item The distribution of numeric attributes like speed and and distance to other actors can be visualized for the set of all traffic scenes which are subgraph isomorphic to an archetype.
    \item It can be cross tabulated, which combinations of archetypes are jointly present in a traffic scene.
    \item Pass Fail rates or other AV performance metrics can be calculated for the subset of all traffic scenes which are subgraph isomorphic to an archetype.
\end{itemize}

In our case, we defined 18 subgraph archetypes covering common traffic scenarios:
\begin{itemize}
    \item Simple 2-actor patterns (only used if 2 actors are isolated): following, opposite traffic, and neighboring vehicles
    \item Complex 3-actor patterns: lead vehicle with neighbor, platoon formations, opposite traffic configurations, and lane change scenarios (cut-in), with variants at intersections
    \item Complex 4-actor patterns: cut-out scenarios, multi-vehicle platoons, and lead-following with opposite traffic, with intersection variants
    \item Complex 5-actor patterns: combined lead, neighbor, and opposite vehicle scenarios with intersection variants
\end{itemize}

It is posible to define more archetypes or detect them automatically, 
but this is out of scope for this project.

\section{Coverage gap analysis using subgraphs}
\label{chapter:analysing_traffic_scene_with_subgraphs}

This chapter systematically identifies coverage gaps—scenarios present in the Argoverse dataset but absent or underrepresented in the CARLA simulation dataset. These \textit{definition holes} provide an intuitive understanding of what traffic situations are missing from the simulated environment. The analysis establishes a foundation for demonstrating that graph embeddings can capture these structural and parametric differences in larger, more complex graphs.


We employ three complementary approaches to identify definition holes in traffic scene dataset, each testing different aspects of traffic scenario representation. For each approach, we assume there is a dataset \textit{TEST} which is compared to another dataset \textit{REF}.

\begin{enumerate}
    \item \textbf{Structural Coverage Analysis}: Identifies scenario archetypes that are significantly underrepresented or absent in the \textit{TEST} dataset compared to \textit{REF}. This tests whether graph embeddings can detect structural differences in graph topology.
    
    \item \textbf{Parametric Distribution Analysis}: Examines differences in continuous node and edge attributes (e.g., vehicle speeds, path lengths) within the same scenario archetype. 
    
    \item \textbf{Co-occurrence Pattern Analysis}: Analyzes which combinations of scenario archetypes appear together in the same traffic scene. This tests whether embeddings can represent more complex, larger subgraph patterns that emerge from the interaction of multiple archetypes.
\end{enumerate}

These concepts will be applied to CARLA and Argoverse datasets in the application chapter \ref{chapter:application}.
\section{Graph Embeddings for Traffic Scene Analysis}
\label{chapter:implementation_of_graph_embeddings_for_traffic_scene_analysis}

The subgraph isomorphism approach presented in Section~\ref{chapter:create_subgraphs_for_coverage_analysis} provides a bottom-up methodology for understanding complex traffic scenarios. While this approach enables systematic analysis of predefined scenario archetypes, the traffic scene graphs constructed from real-world and simulation data exhibit significantly higher complexity than these simple patterns. This motivates the need for complementary top-down approaches, such as graph embeddings, which we explore in this section.
It describes the usage of graph embeddings to traffic scene graphs. 
Embeddings are a widely used method to translate raw data like images or text into an embedding space in order to be 
able to perform machine learning tasks on them. One well known example of this is the Word2Vec model, which is used to 
translate words into a 
vector space, where the distance between vectors can be used to measure the similarity between 
words (\cite{mikolov2013efficientestimationwordrepresentations}).

In the context of traffic scene graphs, embeddings are used to translate the graph structure into a vector space, 
where the distance between vectors can be used to measure the similarity between traffic scenes.
This is useful for coverage analysis, as it allows to compare traffic scenes among each other.
For example, two traffic scenes can be considered similar if the distance between their embeddings is small. This enables
to search for a most similar simulation scenario given a real world scenario, to identify areas with near duplicates or 
to easily visualize structures in the embedding space, which in the original space of all possible traffic scenes
would not be possible.

Graph neural networks (GNNs) are a class of neural networks that are designed to process graph-structured data and have 
gained a lot of popularity in the last years, see for example (add references).

\begin{figure}[H]
    \centering
    \includegraphics[width=0.8\textwidth]{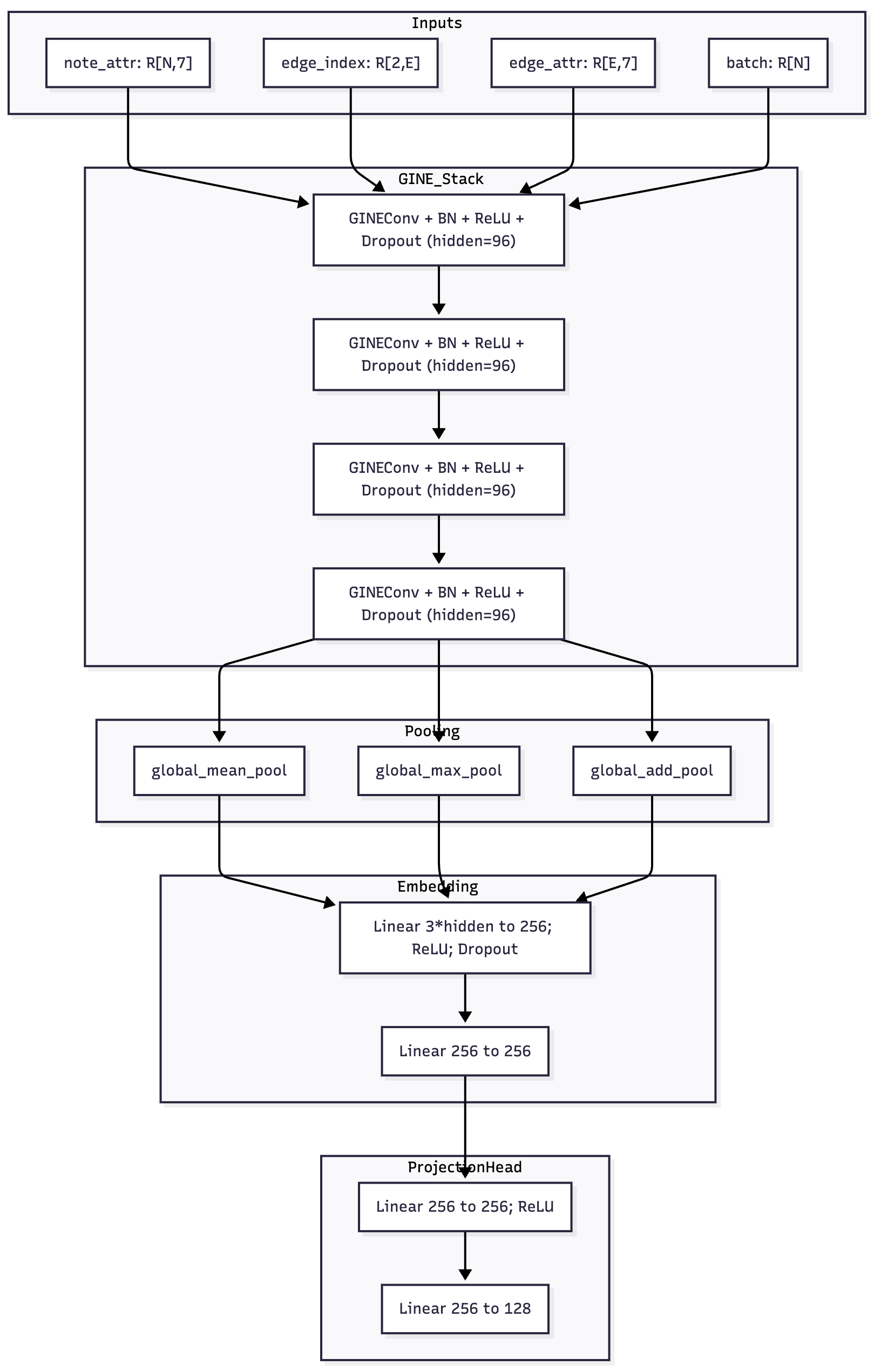}
    \caption{Model architecture for the Graph Isomorphism Network with Edge features (GINE).}
    \label{fig:graph_gine_architecture}
\end{figure}

In this paper, a network architecture using a Graph Isomorphism Network with Edge features (GINE) as described 
in \cite{hu2020strategiespretraininggraphneural} is used to generate embeddings for traffic scene graphs as implemented in the 
pytorch geometric library (\cite{Fey/Lenssen/2019}). While edge-augmented versions of architectures like GraphSAGE and GAT 
now exist \cite{gong2019exploiting}, GINE is chosen specifically because it directly incorporates edge features through 
its aggregation function. This capability is actively utilized in this work: edge type and path length between actors 
are encoded as edge features (see Section \ref{fig:graph_gine_architecture}) and directly influence the message-passing 
process during embedding generation. 

The exact architecture of the model is shown in Figure \ref{fig:graph_gine_architecture}. The features
used are the actor type (as a one-hot encoding with four categories: vehicle, pedestrian, cyclist, motorcycle),
the actor speed (float), if the actor is on an intersection (boolean), and if the actor changed its
lane since the last timestep (boolean) for the nodes.
For the edges, the edge type (as a one-hot encoding with six categories corresponding to different spatial relationships) and the path length (float) between the two nodes are used.

The model has been trained on the CARLA and Argoverse 2.0 datasets using self-supervised contrastive learning 
\cite{chen2020simple}, \cite{hu2020strategiespretraininggraphneural}.

Training employed a self-supervised contrastive objective on mini-batches of 384 graphs.
For each batch, two correlated views of every graph were created through data augmentation:
continuous attributes were perturbed with zero-mean Gaussian noise ($\sigma=0.08$) applied to node
longitudinal speed and edge path length, and edges were randomly dropped with probability 0.1.
A five-layer GINE encoder (hidden width 384) produced graph-level representations via the concatenation of
mean, max, and sum pooling, followed by an embedding MLP (yielding 192-dimensional $\ell_2$-normalized embeddings)
and a projection head. The contrastive loss was a temperature-scaled cross-entropy over in-batch similarities
(cosine similarity of normalized projections, temperature $\tau=0.07$), maximizing agreement of the two views
of the same graph while contrasting against other graphs in the batch. Optimization used AdamW with weight decay
($5 \times 10^{-6}$), learning rate warmup over three epochs, and an exponentially decaying learning rate initialized
at 0.0015 and multiplied by 0.85 across 15 successive training stages.
This setup follows established practice in contrastive pretraining for GNNs \cite{hu2020strategiespretraininggraphneural}
and is implemented using PyTorch Geometric \cite{Fey/Lenssen/2019}.

\section{Application: Analysing traffic scenes with coverage graphs}
\label{chapter:application}

Having defined a graph-based traffic scene representation, we can now analyse the coverage of the system.
Two methodologies are proposed for this purpose:
One is to define archetypes of traffice scenes, and to compare graphs from observed traffic scenes to these archetypes.
The second one is to translate graphs to graph embeddings, and then to compare the embeddings of different sets of traffic scenes.

\subsection{Argoverse 2.0}

Argoverse 2.0 \cite{Argoverse2} is a large-scale dataset for autonomous driving research,
developed by Argo AI. It provides real-world sensor data collected from autonomous vehicle
test fleets operating in six geographically diverse U.S. cities: Austin, Detroit, Miami,
Palo Alto, Pittsburgh, and Washington D.C. The dataset includes high-definition LiDAR point
clouds, ring camera imagery, and detailed vector map information. A key component is the
Motion Forecasting Dataset, which contains 250,000 scenarios of 11 seconds each, featuring
tracked object trajectories for vehicles, pedestrians, cyclists, and other road users.

The dataset has become a standard benchmark in the autonomous driving community, particularly
for motion prediction and trajectory forecasting tasks. Its rich annotations include object
classifications, track identities across frames, and semantic map information such as lane
boundaries, crosswalks, and traffic signal locations. The diverse geographic coverage ensures
exposure to varying road geometries, traffic patterns, and driving behaviors, making it
suitable for developing and evaluating generalizable autonomous driving algorithms.

For this work, the Motion Forecasting Dataset is used, specifically focusing on the tracked
object trajectories and their spatial relationships to construct traffic scene graphs. As the dataset is quite massive, a subset of 17000 scenes from the train partition is used for the analysis.

\subsection{Carla}

CARLA (Car Learning to Act, \cite{Dosovitskiy17Carla}) is an open-source simulator specifically designed for autonomous driving research and 
development. It provides a highly realistic urban driving environment with 
diverse road layouts, weather conditions, and traffic scenarios. The simulator features a comprehensive 
sensor suite simulation, flexible API for scenario creation, and supports both learning-based 
and traditional autonomous driving approaches. CARLA enables researchers to test and 
validate autonomous vehicle systems in a safe, controllable environment before real-world deployment.

The simulator has gained widespread adoption across both academic and industrial settings. In research, CARLA serves as a standard platform for developing and 
benchmarking autonomous driving algorithms, including reinforcement learning approaches for vehicle control and sensor fusion 
techniques \cite{codevilla2019exploringlimitationsbehaviorcloning}. Industry applications include 
virtual testing of production autonomous vehicle systems, scenario-based validation pipelines, and integration 
with hardware-in-the-loop testing frameworks \cite{jaeger2023hiddenbiasesendtoenddriving}. CARLA 
is also extensively used in autonomous driving competitions and challenges, providing a common evaluation 
environment for comparing different approaches across research groups worldwide.

\begin{figure}[H]
\centering
\includegraphics[width=0.8\textwidth]{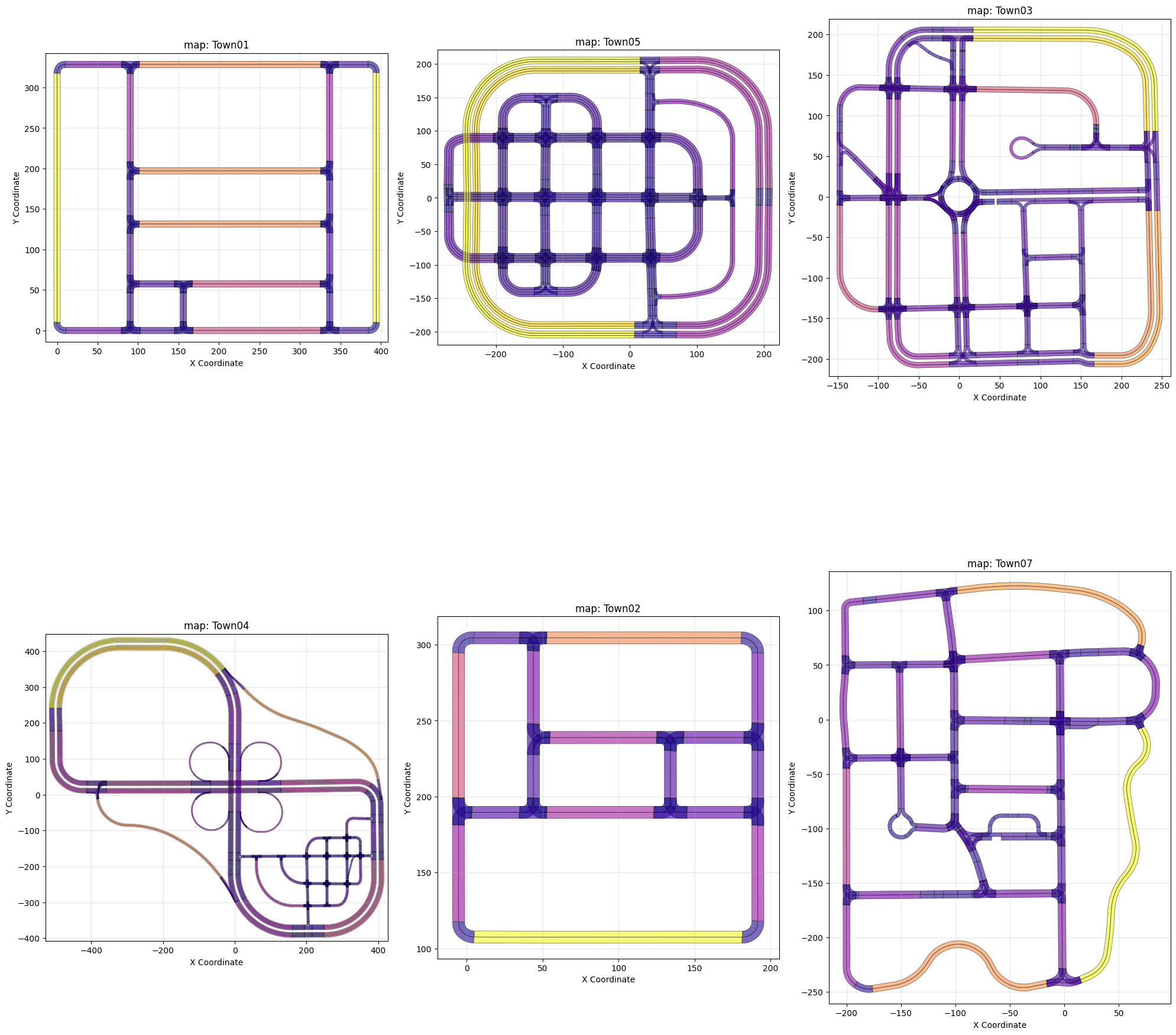}
\caption{Overview of CARLA maps used in the simulation study: Town01, Town02, Town03, Town04, Town05, and Town07. These maps provide diverse urban driving environments with varying road layouts, intersections, and traffic patterns.}
\label{fig:carla_maps}
\end{figure}

Here, Carla version $0.9.15$ is used. The CARLA version $0.10.0$ is not used, because it had only 2 maps and
Mine\_1 (which is not really normal roads) at the start of this project.
Specifically, the following maps were used: Town01, Town02, Town03, Town04, Town05 and Town07. Plots
of these maps are shown in Figure \ref{fig:carla_maps}.

A data generation script implements some behavior control mechanisms to create 
diverse and realistic traffic scenarios. Multiple vehicle types including trucks, motorcycles, 
and regular cars are spawned with varying probabilities, each exhibiting different behavioral
characteristics such as speed preferences, following distances, and lane-changing tendencies. 
The script incorporates dynamic behavior modifications during simulation, including random slowdowns, 
periodic behavior changes, and adaptive responses to traffic conditions, resulting in rich and 
varied traffic scene data across multiple CARLA maps and simulation iterations. The simulation runs
have between 20 and 60 vehicles each.

The resulting data consists of 2050 scenes with 11 seconds of simulation time each.

\subsection{Subgraph Isomorphism Coverage Analysis}

We apply the subgraph isomorphism approach to both the  CARLA and the Argoverse data to identify coverage scenarios.
The archetypes used here are defined manually and described in table \ref{tab:subgraph_archetypes}. The chosen archetypes are typical traffic situations like e.g. lead vehicle situations, lane change situations or different combinations of opposite direction vehicles.

\begin{table}[H]
\centering
\caption{Overview of all subgraph archetypes with their structural properties. Edge types: \textit{fl} = following\_lead, \textit{lv} = leading\_vehicle, \textit{nv} = neighbor\_vehicle, \textit{ov} = opposite\_vehicle.}
\label{tab:subgraph_archetypes}
\begin{tabular}{llccl}
\toprule
\textbf{Group} & \textbf{Archetype} & \textbf{Actors} & \textbf{Edges} & \textbf{Edge Types} \\
\midrule
\multirow{3}{*}{Simple (2-actor)}
  & Simple Following           & 2 & 2 & fl, lv \\
  & Simple Opposite            & 2 & 2 & ov \\
  & Simple Neighbor            & 2 & 2 & nv \\
\midrule
\multirow{8}{*}{Complex (3-actor)}
  & Lead + Neighbor (intersection)       & 3 & 4 & fl, lv, nv \\
  & Cut-in                               & 3 & 4 & fl, lv \\
  & Cut-in (intersection)                & 3 & 4 & fl, lv \\
  & Platoon (intersection)               & 3 & 4 & fl, lv \\
  & Opposite Traffic (intersection)      & 3 & 4 & fl, lv, ov \\
  & Lead + Neighbor at Intersection      & 3 & 4 & fl, lv, nv \\
  & Triple Opposite (intersection)       & 3 & 4 & ov \\
  & Lead + Following in Back             & 3 & 4 & fl, lv \\
\midrule
\multirow{5}{*}{Complex (4-actor)}
  & Lead + Neighbor                      & 3 & 4 & fl, lv, nv \\
  & Cut-out                              & 4 & 6 & fl, lv, nv \\
  & Cut-out (intersection)               & 4 & 6 & fl, lv, nv \\
  & 4-Vehicle Platoon (intersection)     & 4 & 6 & fl, lv \\
  & 4-Vehicle Opposite (intersection)    & 4 & 6 & fl, lv, ov \\
\midrule
\multirow{2}{*}{Complex (5-actor)}
  & Lead + Neighbor + Opposite           & 5 & 8 & fl, lv, nv, ov \\
  & Lead + Neighbor + Opposite (inters.) & 5 & 8 & fl, lv, nv, ov \\
\bottomrule
\end{tabular}
\end{table}

\begin{figure}[H]
    \centering
    \includegraphics[width=0.85\textwidth]{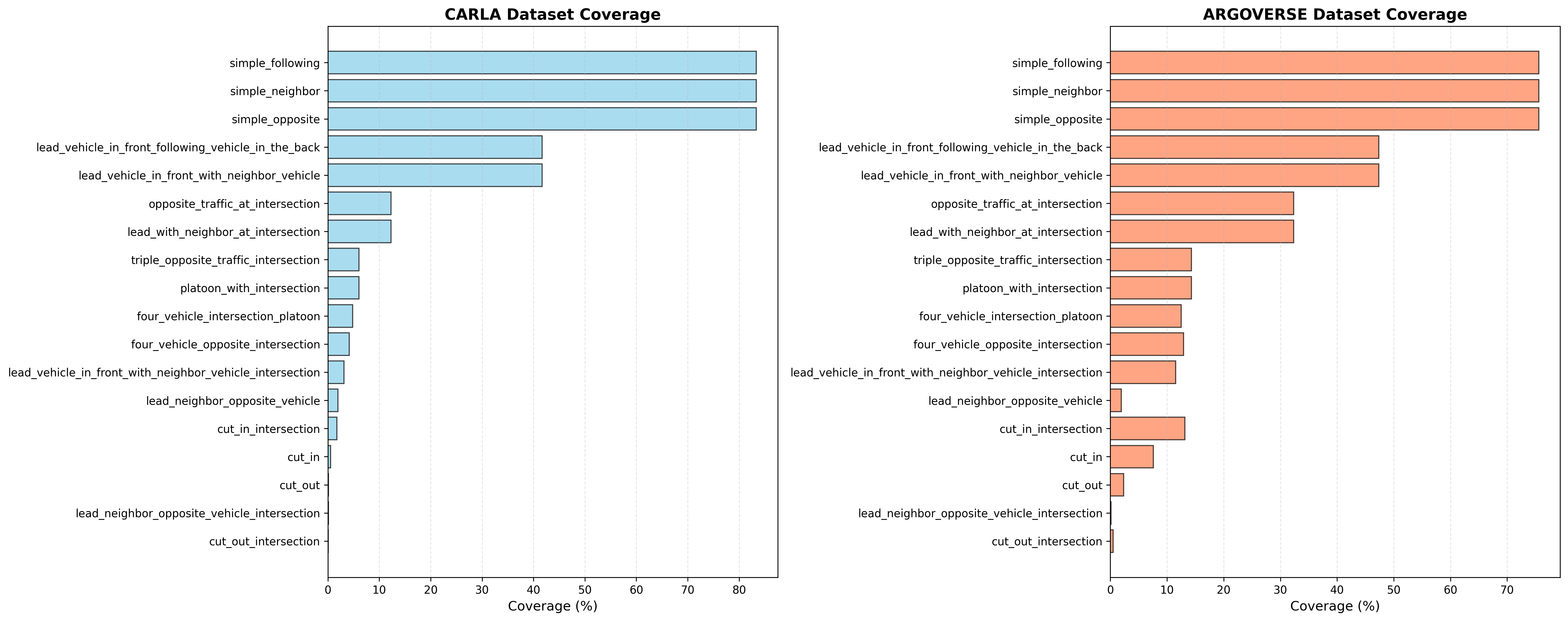}
    \caption{Coverage barcharts for the manually defined coverage scenarios for CARLA and Argoverse.}
    \label{fig:subgraph_isomorphism_coverage_barcharts}
\end{figure}

\begin{figure}[H]
    \centering
    \begin{subfigure}[b]{0.48\textwidth}
        \centering
        \includegraphics[width=\textwidth]{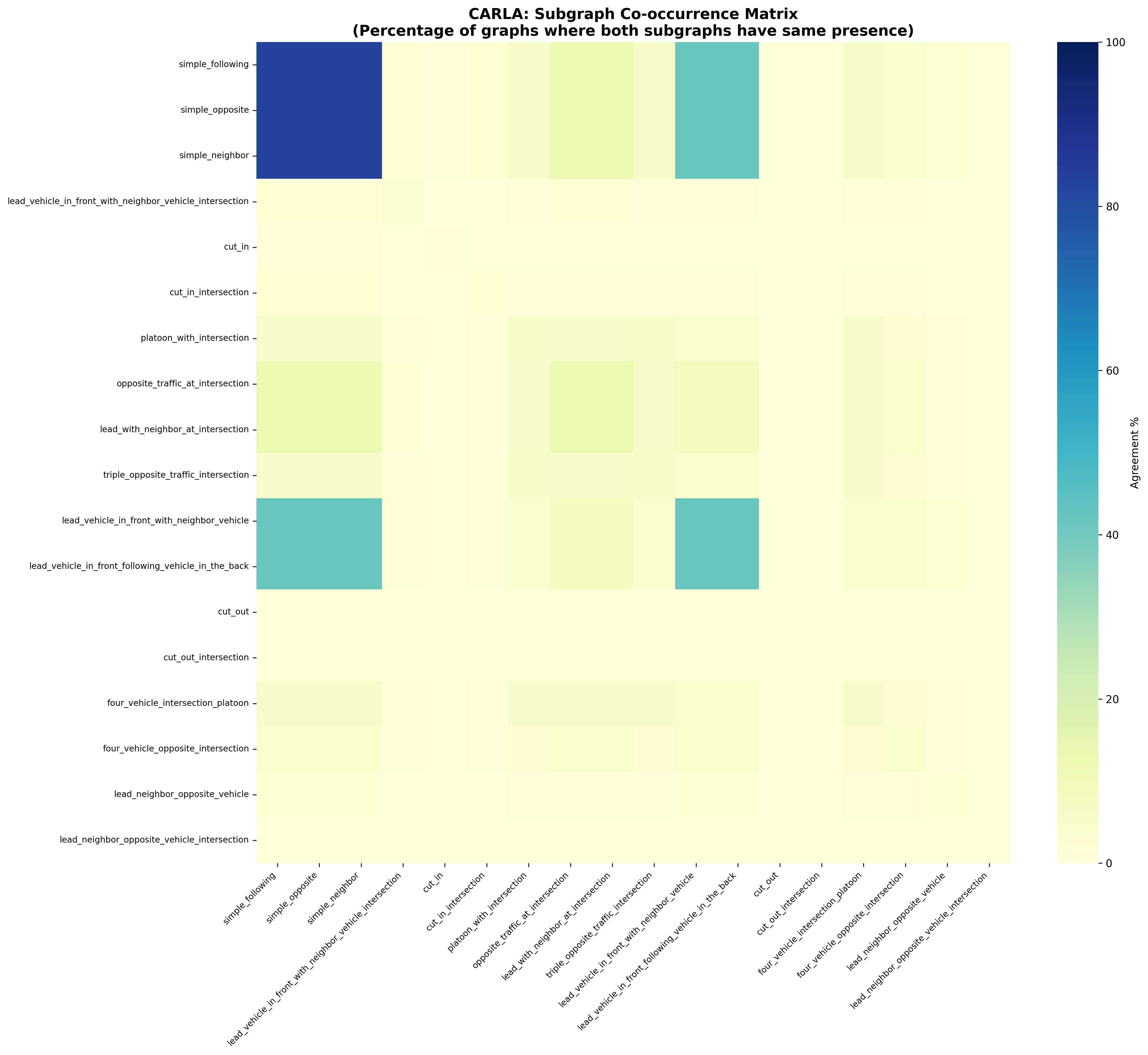}
        \caption{Co-occurrence matrix for CARLA.}  
        \label{fig:carla_cooccurrence_matrix}
    \end{subfigure}
    \hfill
    \begin{subfigure}[b]{0.48\textwidth}
        \centering
        \includegraphics[width=\textwidth]{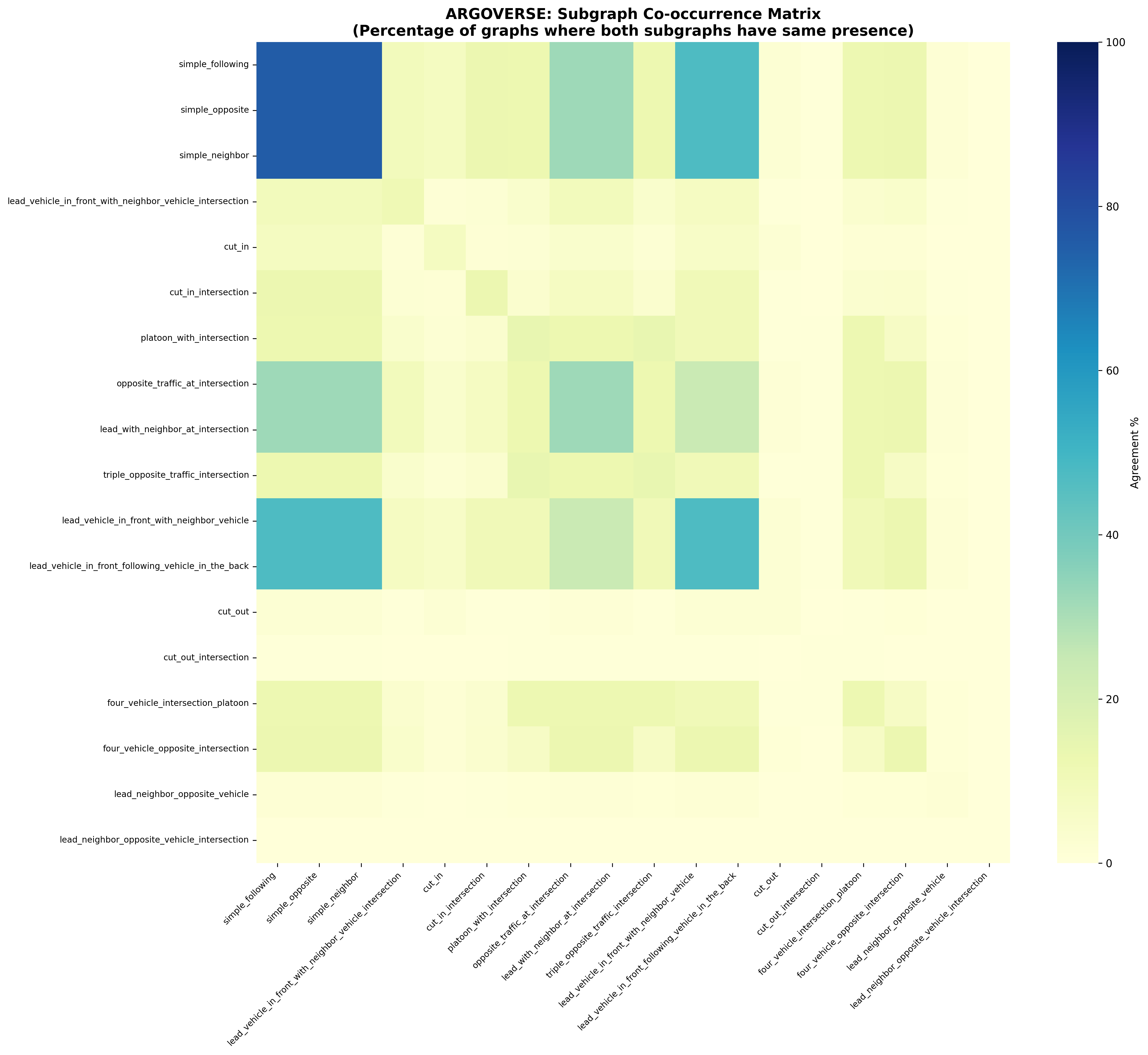}
        \caption{Co-occurrence matrix for Argoverse.}  
        \label{fig:argo_cooccurrence_matrix}
    \end{subfigure}
    \caption{Co-occurrence matrices showing the percentage of scenarios where pairs of subgraph patterns appear together in the same traffic scene for CARLA and Argoverse datasets.}
    \label{fig:cooccurrence_matrices}
\end{figure}

\begin{figure}[H]
    \centering
    \includegraphics[width=0.85\textwidth]{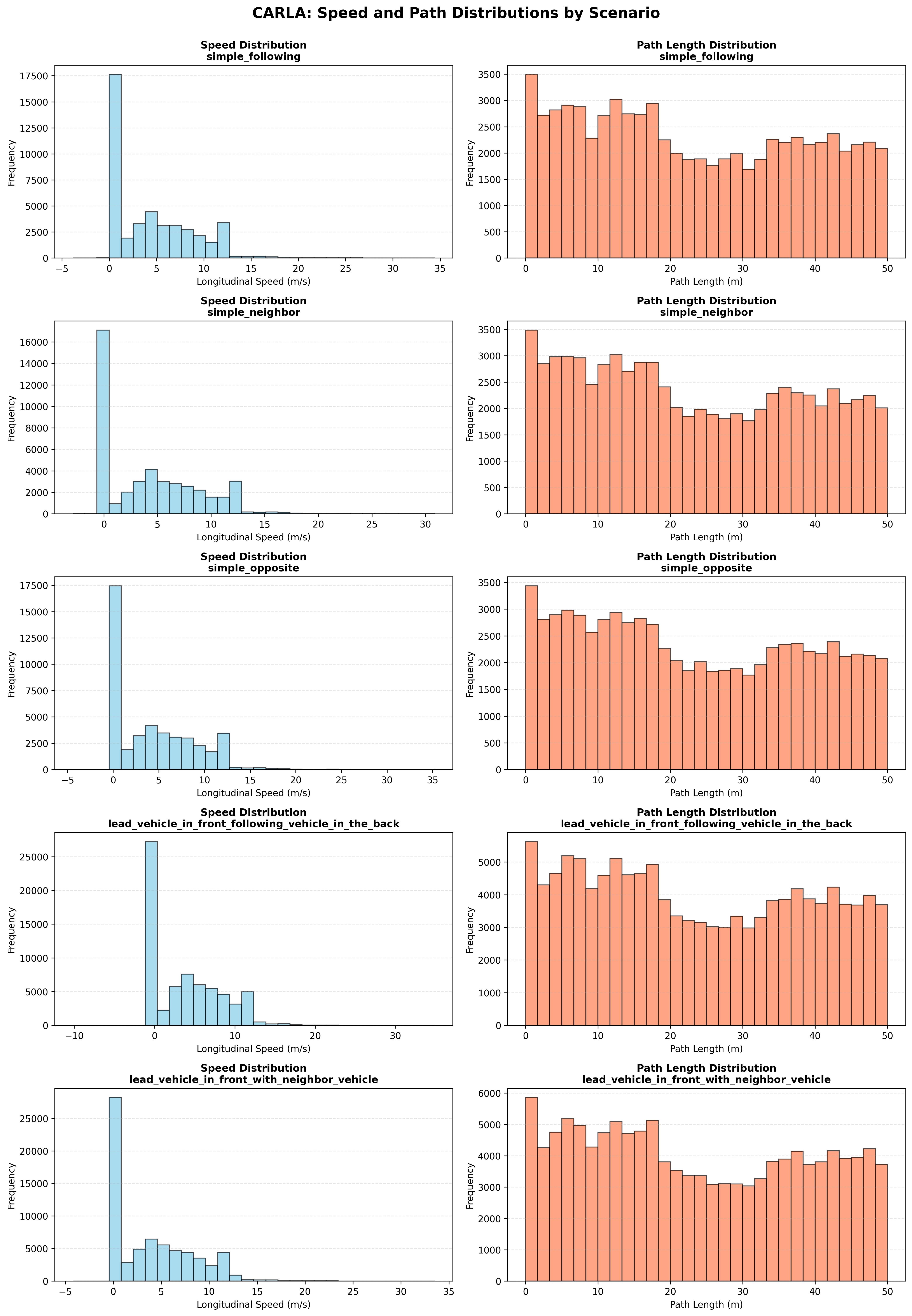}
    \caption{Speed and path length distributions for the top 5 most common scenario archetypes in CARLA. Each row shows a different scenario with speed (left) and path length (right) distributions.}
    \label{fig:carla_speed_path_distributions}
\end{figure}

\begin{figure}[H]
    \centering
    \includegraphics[width=0.85\textwidth]{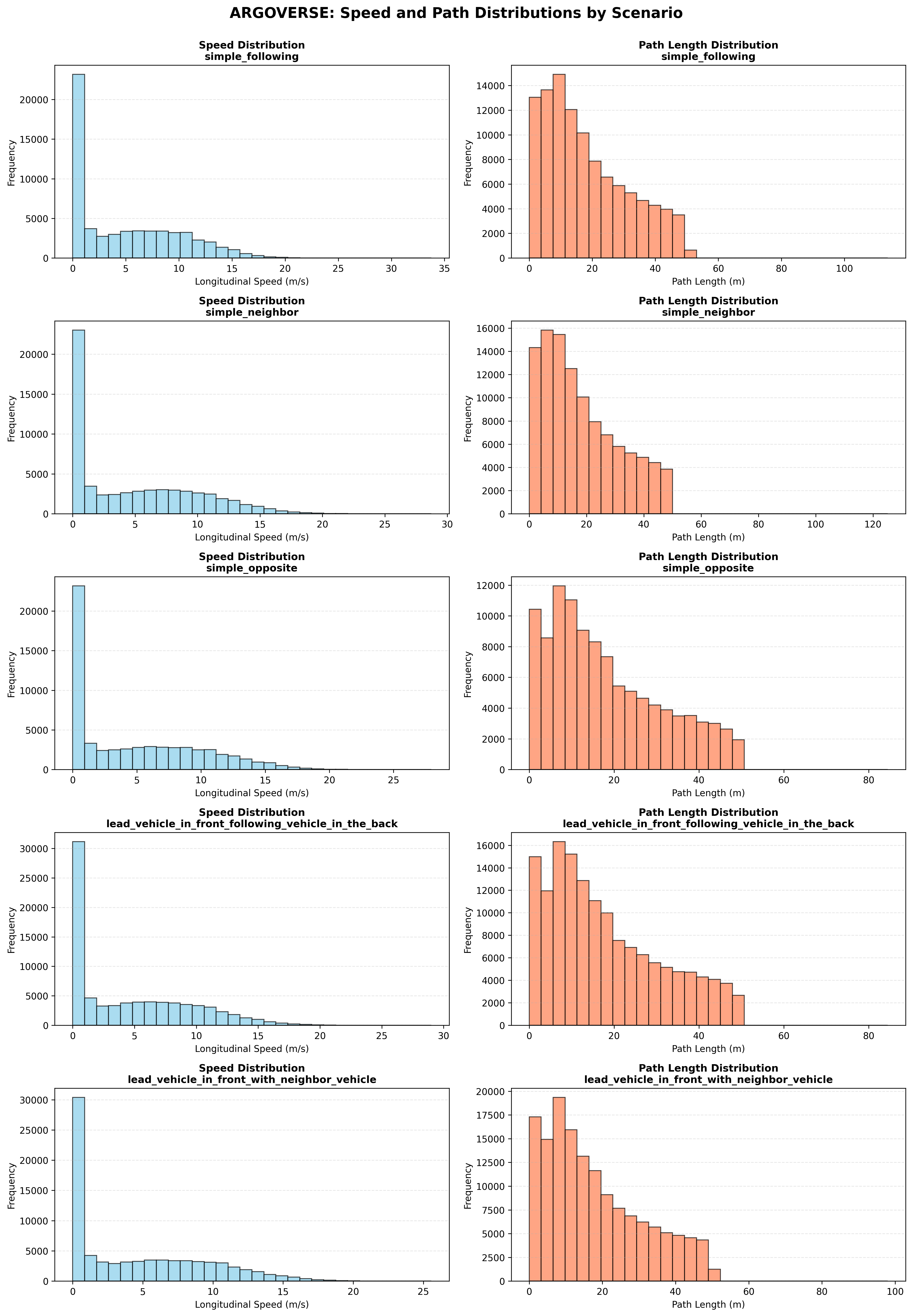}
    \caption{Speed and path length distributions for the top 5 most common scenario archetypes in Argoverse. Each row shows a different scenario with speed (left) and path length (right) distributions.}
    \label{fig:argo_speed_path_distributions}
\end{figure}

Also, information like which traffic actors are on an intersection are incorporated into the archetypes. Overall, the node coverage analysis across Argoverse and CARLA datasets achieved 62\% overall coverage. 
The results are shown in Figure \ref{fig:subgraph_isomorphism_coverage_barcharts} to Figure \ref{fig:argo_speed_path_distributions}.
In Figure \ref{fig:subgraph_isomorphism_coverage_barcharts} there is a clear signal that for both datasets the coverage is not uniform per the different archetypes.
Also, the distribution for the Carla dataset is not close to the Argoverse distribution, indicating that the CARLA data is not representative for the Argoverse data.
Even worse, the Carla dataset is nearly completely missing out e.g. on the cut\_out\_intersection archetype, clearly indicating a  rather randomly generated Carla dataset.
A next step of analysis is to check, which archetypes are occuring simultaneously in a traffic scene. 
Figure \ref{fig:cooccurrence_matrices} shows the agreement matrix for the manually defined coverage scenarios for CARLA and Argoverse.
The heatmaps show the percentage of agreement between the manually defined coverage scenarios for CARLA and Argoverse.


A last example of how to use the assignment of archetypes to traffic scenes is to check the parameter distribution for the speed and path length of the traffic scenes.
Figure \ref{fig:carla_speed_path_distributions} and Figure \ref{fig:argo_speed_path_distributions} show the parameter distribution for the speed and path length of the traffic scenes for CARLA and Argoverse.
The plots show that the distribution for the CARLA data differs from the Argoverse distribution, indicating that the CARLA data is not representative for the Argoverse data.
Given the assignment of archetypes to traffic scenes based on the actors can be used in further analysis not done here, e.g. to check more divers parameters and distributions.


\subsection{Coverage gap analysis using subgraphs}

\subsubsection{Structural Coverage Analysis}

\begin{figure}[H]
    \centering
    \includegraphics[width=0.95\textwidth]{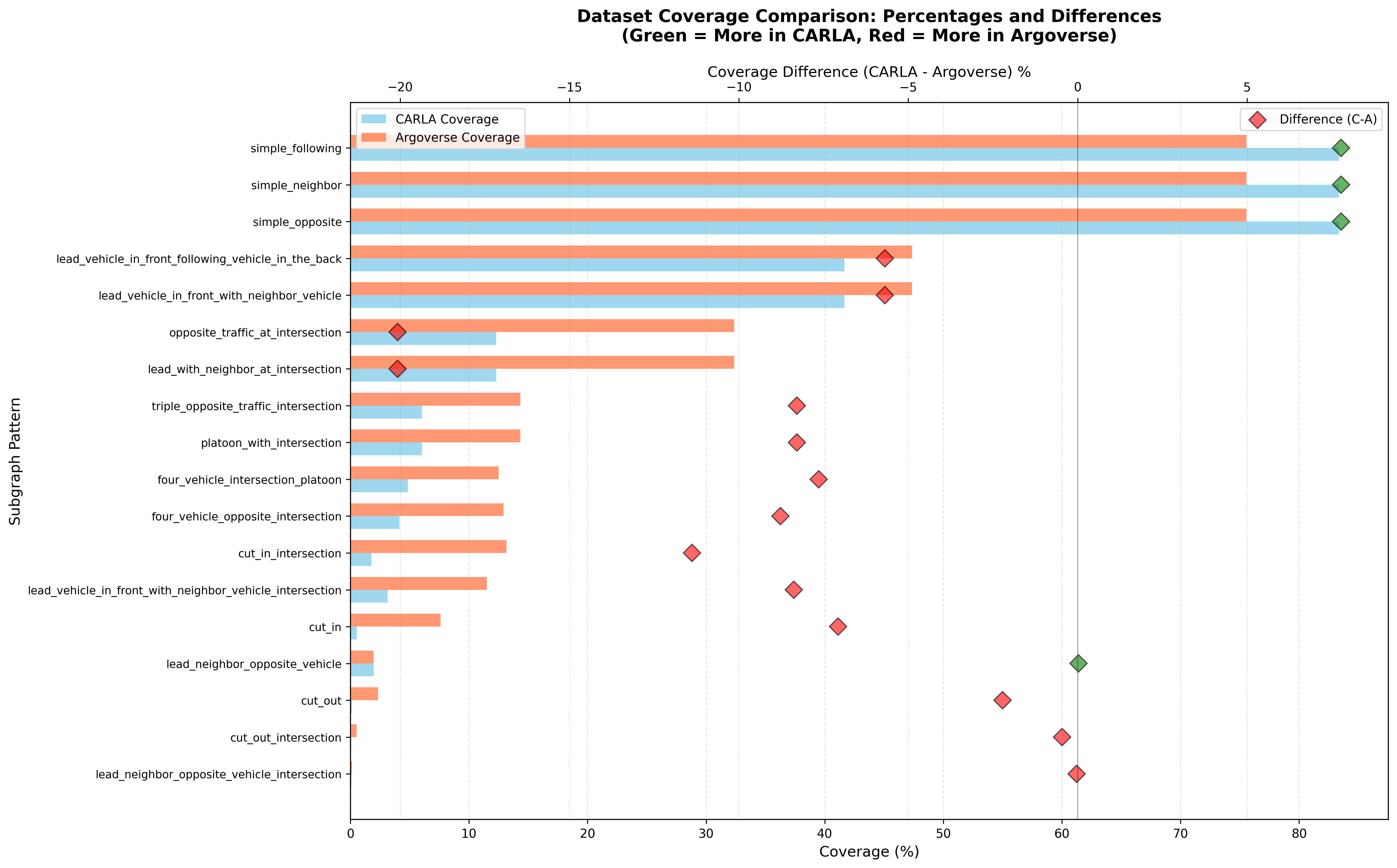}
    \caption{Subgraph coverage comparison between CARLA and Argoverse datasets. The bar chart shows relative coverage percentages for each scenario archetype, while diamond markers indicate the magnitude and direction of coverage differences. Patterns sorted by average coverage across both datasets.}
    \label{fig:coverage_comparison}
\end{figure}

Figure~\ref{fig:coverage_comparison} presents a comprehensive comparison of subgraph archetype coverage across the two datasets. The dual-axis visualization shows both the relative coverage percentages (left y-axis) and their differences (right x-axis, shown as diamond markers). Green markers indicate scenarios more prevalent in CARLA, while red markers highlight scenarios more common in Argoverse.

The analysis reveals several structural definition holes in the CARLA dataset. Most notably, complex multi-actor scenarios such as intersection-based patterns show significant underrepresentation. For instance, the \newline \texttt{complex\_4actor\_cut\_out\_intersection}  pattern appears in 8.2\% of Argoverse scenes but only 0.3\% of CARLA scenes, representing a coverage gap of 7.9 percentage points. Similarly, \newline \texttt{complex\_3actor\_opposite\_traffic\_intersection} exhibits a gap of 6.5 percentage points (9.1\% in Argoverse vs. 2.6\% in CARLA).

Conversely, CARLA demonstrates higher coverage for simpler highway scenarios. The \texttt{simple\_2actor\_following} pattern appears in 42.3\% of CARLA scenes compared to 31.8\% in Argoverse, suggesting that the CARLA simulation emphasizes highway driving conditions over complex urban intersections.

\subsubsection{Parametric Distribution Analysis}

Beyond structural differences, we analyze how continuous node and edge attributes differ within the same scenario archetype. Even when a scenario archetype is present in both datasets, the distributions of speed and path length can reveal definition holes at the parameter level.

\subsubsection{Role-Specific Speed Distribution Holes}

\begin{figure}[H]
    \centering
    \includegraphics[width=0.95\textwidth]{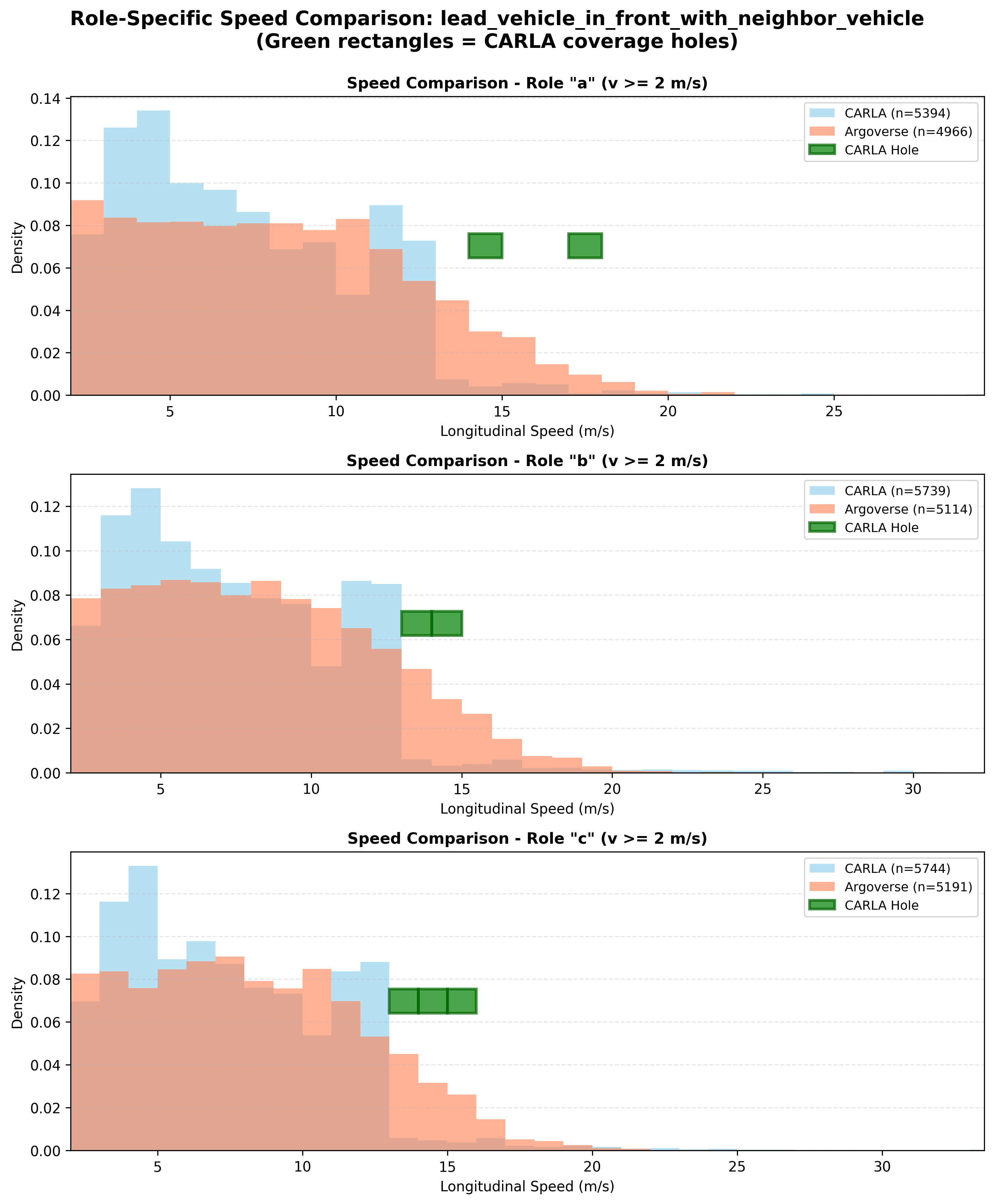}
    \caption{Role-specific speed distribution comparison for the lead vehicle with neighbor scenario. Green rectangles mark definition holes—speed ranges where Argoverse has significant density ($\geq$ 0.5\%) but CARLA has less than 15\% of that density. Each row represents a different actor role within the scenario (e.g., lead vehicle, ego vehicle, neighbor vehicle).}
    \label{fig:role_speed_holes}
\end{figure}

To investigate parametric differences in greater depth, we analyze speed distributions at the actor-role level within each scenario archetype. Figure~\ref{fig:role_speed_holes} presents an exemplary analysis for the \texttt{lead\_vehicle\_in\_front\_with\_neighbor\_vehicle} scenario, which represents a common highway situation with lane change potential.

To systematically identify definition holes, we employ two threshold criteria: (1) Argoverse must show significant presence in a speed bin with a density of at least 0.5\%, ensuring the speed range is meaningfully represented in real-world data, and (2) CARLA's density in that bin must be less than 15\% of Argoverse's density, indicating substantial underrepresentation rather than minor variations. Speed bins meeting both criteria are marked as definition holes.

The green rectangles in Figure~\ref{fig:role_speed_holes} highlight multiple speed distribution holes across different actor roles in this scenario. These gaps indicate that CARLA underrepresents certain speed ranges common in real-world highway driving, particularly at moderate to higher speeds where Argoverse shows consistent density but CARLA has minimal coverage.

Similar patterns of parametric definition holes are observed across other scenario archetypes. For instance, the \texttt{lead\_vehicle\_in\_front\_with\_neighbor\_vehicle} scenario exhibits multiple speed distribution holes for the neighbor vehicle role, particularly in the 12-15 m/s range, suggesting insufficient coverage of moderate-speed lane change scenarios in CARLA.

This role-specific analysis demonstrates that definition holes can exist at the parameter level even when the structural scenario archetype is present in both datasets. Such parametric differences encode information in the node attributes that should be detectable by graph embedding methods.

\subsection{Co-occurrence Pattern Analysis}

Traffic scenes in the real world often contain multiple scenario archetypes simultaneously. For example, a vehicle might be following a lead vehicle while also navigating an intersection with opposite traffic. To capture these more complex patterns, we analyze co-occurrence matrices that quantify how frequently pairs of archetypes appear together in the same traffic scene.

\begin{figure}[H]
    \centering
    \includegraphics[width=0.95\textwidth]{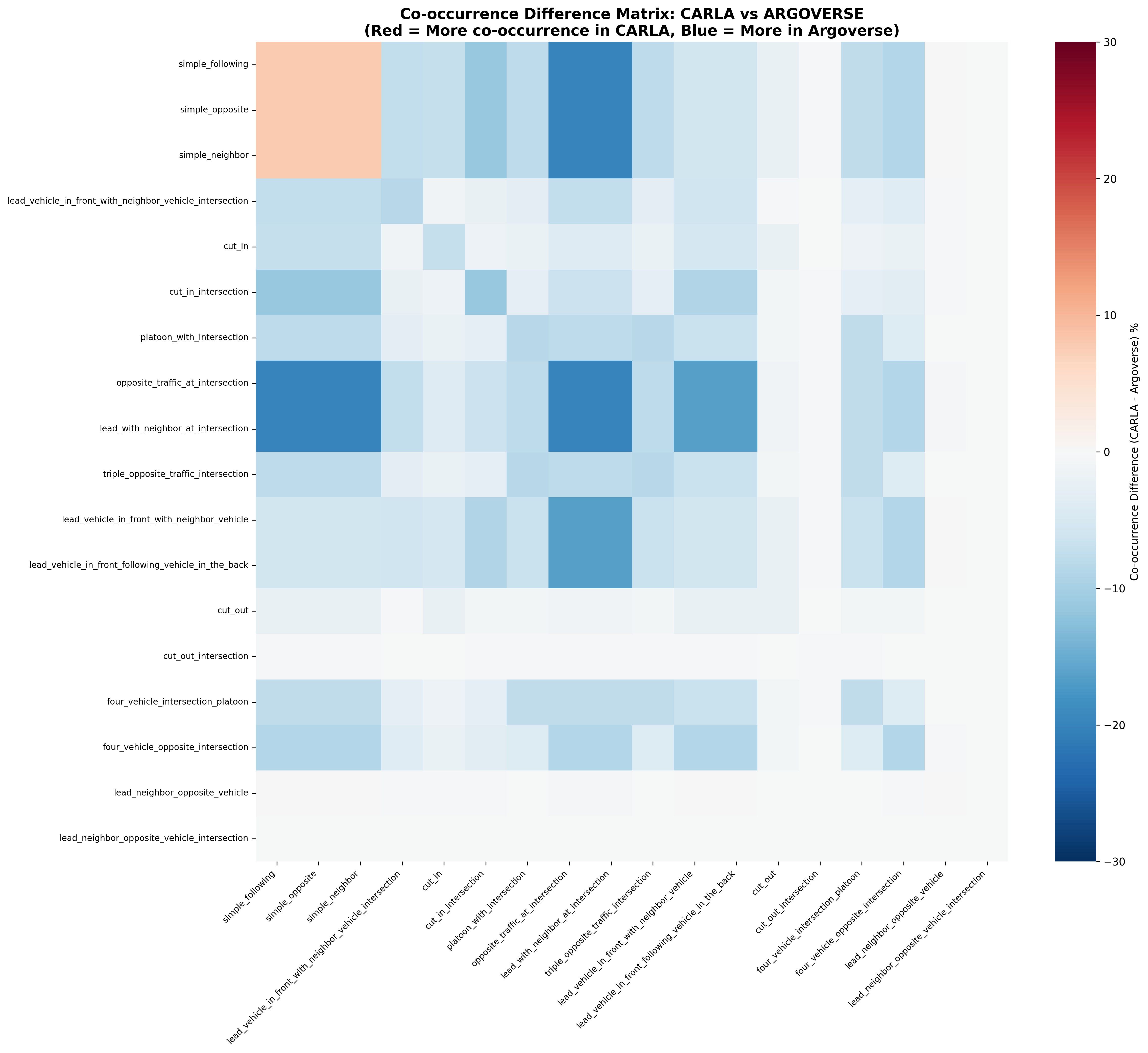}
    \caption{Co-occurrence difference matrix showing where pairs of scenario archetypes appear together significantly more in Argoverse (red) or CARLA (blue). Values represent percentage point differences in joint occurrence rates.}
    \label{fig:cooccurrence_diff}
\end{figure}

Figure~\ref{fig:cooccurrence_diff} visualizes the differences in co-occurrence rates between the two datasets. Each cell $(i,j)$ represents the percentage point difference $P_{\text{Argo}}(i \cap j) - P_{\text{CARLA}}(i \cap j)$, where $P(i \cap j)$ denotes the probability that both archetypes $i$ and $j$ are present in the same scene.

The most significant co-occurrence holes are:

\begin{itemize}
    \item \textbf{Following + Overtaking}: This combination appears in 15.2\% of Argoverse scenes but only 1.8\% of CARLA scenes (gap: 13.4 percentage points), indicating that CARLA rarely simulates scenarios where vehicles simultaneously follow and overtake other traffic.
    
    \item \textbf{Merging + 3-Chain}: Present in 12.7\% of Argoverse scenes vs. 2.1\% in CARLA (gap: 10.6 percentage points), reflecting underrepresentation of complex multi-vehicle merging maneuvers in CARLA.
    
    \item \textbf{Crossing + 4-Intersection}: Appears in 11.8\% of Argoverse scenes vs. 1.5\% in CARLA (gap: 10.3 percentage points), consistent with CARLA's general underrepresentation of complex intersection scenarios.
\end{itemize}

These co-occurrence holes indicate that CARLA not only lacks certain individual scenario archetypes but also fails to generate realistic combinations of archetypes that commonly appear together in real-world driving. Since these patterns involve larger, more complex subgraphs that span multiple archetypes, they provide a test case for whether graph embeddings can capture information at a higher level of abstraction than simple structural or parametric differences.

\subsubsection{coverage gap analysis summary}

This three-tiered analysis systematically identifies definition holes in the CARLA dataset across structural, parametric, and co-occurrence dimensions. These findings serve a dual purpose:

\begin{enumerate}
    \item They provide actionable insights for improving simulation-based testing by highlighting which traffic scenarios need better representation.
    
    \item They establish ground truth for evaluating graph embedding methods in subsequent sections. If embeddings successfully encode the relevant information, they should enable detection of these holes without explicit subgraph isomorphism checks, thereby providing a scalable top-down approach to coverage analysis.
\end{enumerate}

The structured, bottom-up approach presented here demonstrates that subgraph isomorphism can effectively characterize traffic scene coverage for predefined archetypes. However, the manual definition of archetypes and the computational cost of isomorphism checking for large scenario collections motivate the graph embedding approach explored in the following chapter.

\subsection{Graph Embeddings Coverage Analysis}

\begin{figure}[H]
    \centering
    \includegraphics[width=0.8\textwidth]{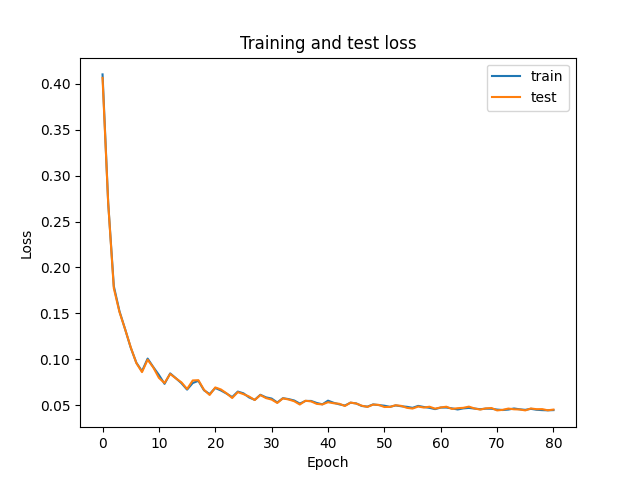}
    \caption{Training and test loss for the graph embeddings model trained jointly on CARLA and Argoverse 2.0 data.}
    \label{fig:train_test_graph_embeddings_loss_plot}
\end{figure}

\begin{figure}[H]
    \centering
    \includegraphics[width=0.95\textwidth]{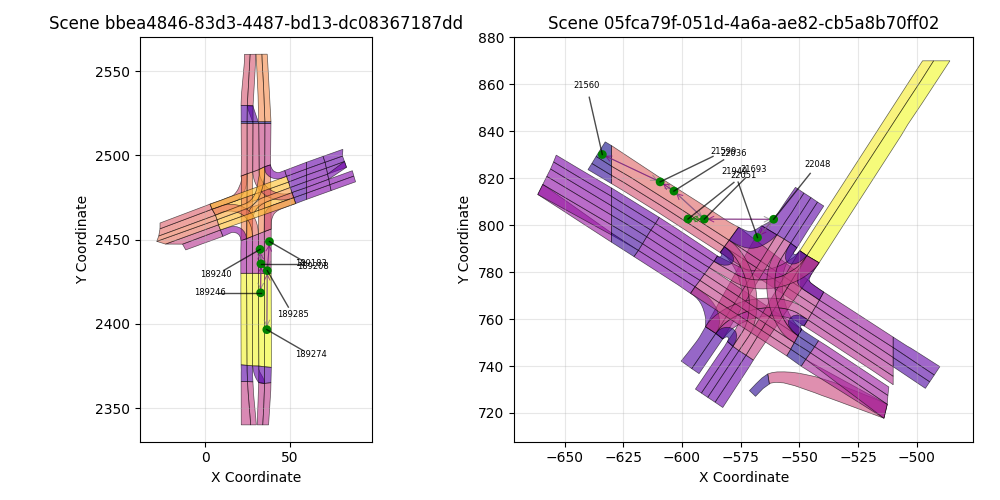}
    
    \includegraphics[width=0.95\textwidth]{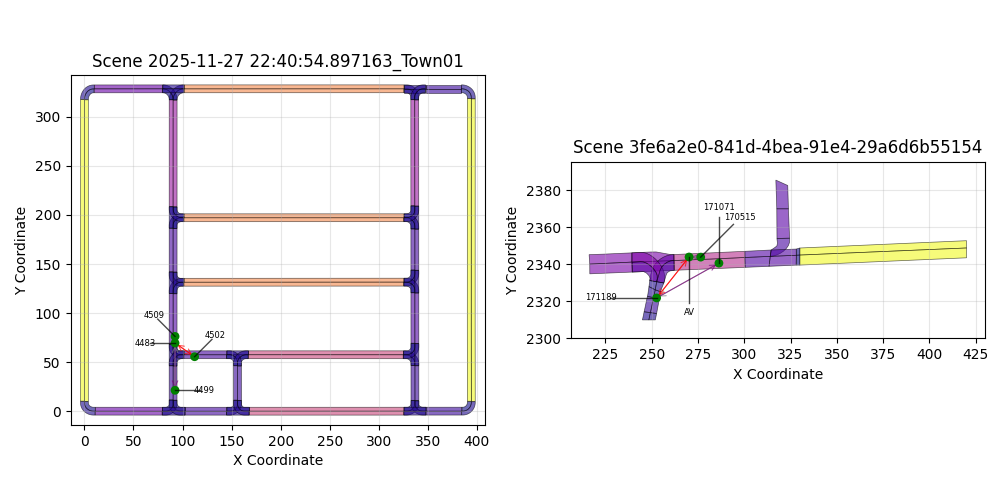}
    
    \includegraphics[width=0.95\textwidth]{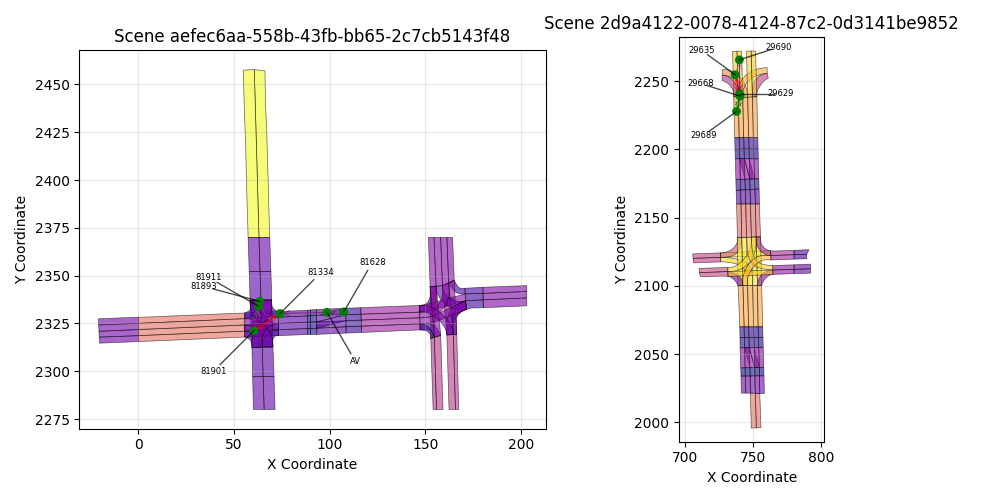}
    
    \includegraphics[width=0.95\textwidth]{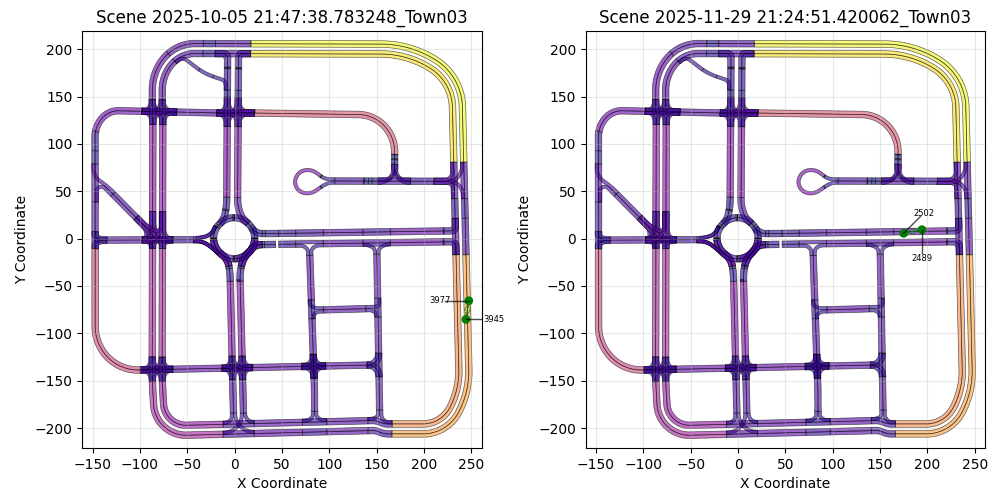}
    
    \includegraphics[width=0.95\textwidth]{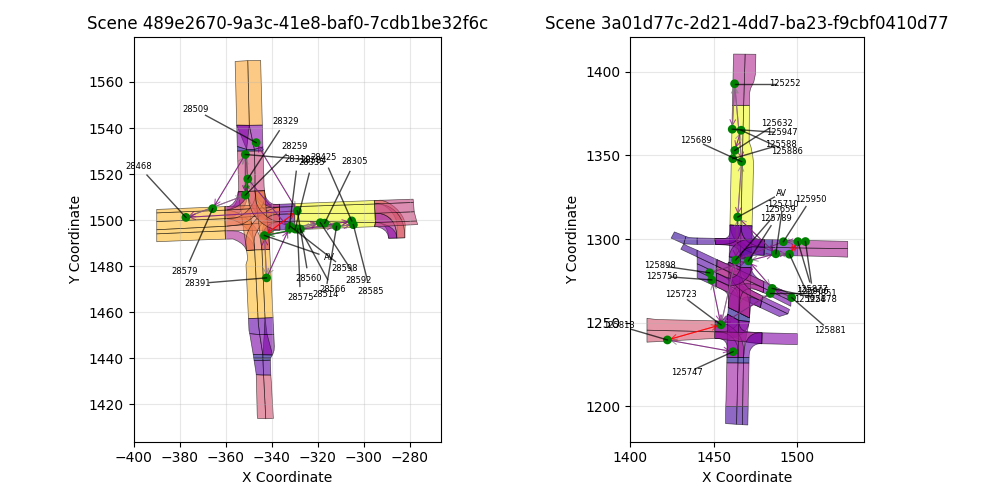}
    \caption{Plausibility checks showing graph embedding comparisons. For randomly sampled scenarios, the most similar scenarios based on embedding distance (both Euclidean and cosine) are visualized, including comparisons between CARLA and Argoverse scenarios.}
    \label{fig:plausibility_checks}
\end{figure}

The resulting embeddings have been analysed in a number of ways. For plausibility checks, for a number of 
randomly samples scenarios, the scenario with 
closest embedding vector (euclidean distance) has been visualized. 
This is shown in Figure \ref{fig:plausibility_checks}. This 
includes comparison between CARLA and Argoverse scenarios.

\begin{figure}[H]
    \centering
    \includegraphics[width=0.8\textwidth]{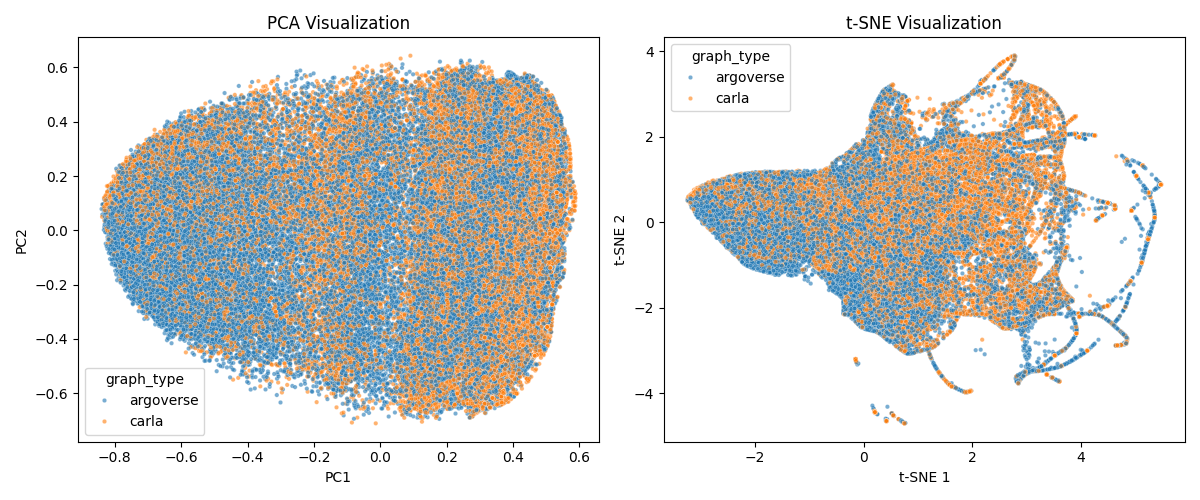}
    \caption{PCA and t-SNE visualization of the embedding space for Carla and Argoverse 2.0 scenarios.}
    \label{fig:embedding_space}
\end{figure}
    
As a next step, the embedding space has been analysed using dimension reduction techniques, specifically PCA and t-SNE. This is shown in Figure \ref{fig:embedding_space}. This can be done in a number of 
different flavors, like for example:

\begin{itemize}
    \item distinguishing between CARLA and Argoverse scenarios by a color coding, 
    \item visualizing just the CARLA scenarios and color code them by the map, in order to see if the maps deliver different types of scenarios,
    \item by calculating a density distribution of the embedding space for the Argoverse scenarios, and to check if for a regions with a density 
    about a certain threshold, there exist CARLA scenarios, i.e. do a coverage analysis in the embedded space
    \item do the same vice versa, i.e. to check for relevance of CARLA scenarios.
\end{itemize}

As the embedding space is a normal metric space, representing scenarios across a large range of different driving situations, these embeddings can be used also for 
many other tasks, like for example clustering, anomaly detection, similarity search, etc.

And, as the main task considered here is coverage analysis, the embeddings can be used to check if a target distribution is met by a test distribution in the embedded space. 
Specifically, considering the Argoverse 2.0 scenarios as the target distribution, and the CARLA scenarios as the test distribution, the embeddings can be used to check 
if the CARLA scenarios cover the Argoverse 2.0 scenarios in the embedded space.

\section{Summary}

This paper presented a graph-based framework for coverage analysis in autonomous driving that represents traffic scenes as hierarchical graphs combining map topology with actor relationships.
The framework employs a two-phase graph construction algorithm to systematically capture spatial relationships between traffic participants, encoding leading, following, neighboring, and opposing configurations as typed edges in a directed multigraph structure.

Two complementary approaches for coverage analysis were developed and evaluated.
The subgraph isomorphism approach utilizes a set of manually defined archetype graphs representing common driving scenarios.
This method provides interpretable coverage metrics by identifying which specific traffic patterns are present in a given scene.
The graph embedding approach leverages Graph Isomorphism Networks with Edge features trained via self-supervised contrastive learning to project traffic scenes into a continuous vector space.
This representation enables similarity-based coverage assessment, clustering of related scenarios, and anomaly detection for identifying underrepresented traffic situations with arbitrary number of actors.

Both approaches were validated on real-world data from the Argoverse 2.0 dataset and synthetic data from the CARLA simulator, as presented in Section~\ref{chapter:application}.
The experimental results demonstrate that the framework successfully captures meaningful traffic scene structure and reveals notable differences in scenario distributions between the two datasets.

The proposed approach offers several advantages over traditional coverage analysis methods.
Most significantly, it scales efficiently to diverse traffic scenarios without requiring scenario-specific handling rules.
In contrast to approaches such as TNO Streetwise \cite{tno_streetwise}, where each scenario type must be manually defined and processed individually, the presented framework only requires the definition of actor graph construction rules and, optionally, archetype graphs.
Furthermore, the approach naturally accommodates varying numbers of actors in a scene without excluding participants or focusing on arbitrary subsets.

Future work will extend the framework in two directions.
First, the actor graph representation will be enhanced to incorporate temporal information by including multiple time steps within a single graph structure.
Second, methods for automatically extracting archetype graphs from real-world traffic observation data will be investigated, reducing the reliance on manual archetype definition.

The source code for this research is publicly available at \url{https://github.com/tmuehlen80/graph_coverage}.

\bibliographystyle{splncs04}
\bibliography{lit}

\end{document}